\documentclass[final]{siamltex}
\usepackage{euscript,amsmath,amssymb,amsfonts,graphicx,bm,cite,appendix}

 \usepackage[colorlinks=true]{hyperref}
 \hypersetup{urlcolor=blue, citecolor=red}

\newcommand{\R}{{\mathbb R}}
\newcommand{\mc}{\mathcal}
\newcommand{\pd}{\partial}

\newcommand{\e}{{\rm e}}

\newcommand{\ve}{\varepsilon}

\newcommand{\erf}{{\rm erf}}
\renewcommand{\d}{{\rm d}}

\title{Ghosts of bump attractors in stochastic neural fields: Bottlenecks and extinction}

\author{Zachary P. Kilpatrick\thanks{Department of Mathematics, University of Houston, Houston, Texas 77204, USA ({\tt zpkilpat@math.uh.edu}). This author is supported by NSF grants (DMS-1311755).}}

\date{\today}

\begin{document}

\maketitle
\newcommand{\slugmaster}{%
\slugger{MMedia}{xxxx}{xx}{x}}

\begin{abstract}We study the effects of additive noise on stationary bump solutions to spatially extended neural fields near a saddle-node bifurcation. The integral terms of these evolution equations have a weight kernel describing synaptic interactions between neurons at different locations of the network. Excited regions of the neural field correspond to parts of the domain whose fraction of active neurons exceeds a sharp threshold of a firing rate nonlinearity. For sufficiently low firing threshold, a stable bump coexists with an unstable bump and a homogeneous quiescent state. As the threshold is increased, the stable and unstable branch of bump solutions annihilate in a saddle node bifurcation. Near this criticality, we derive a quadratic amplitude equation that describes the slow evolution of the even mode (bump contractions) as it depends on the distance from the bifurcation. Beyond the bifurcation, bumps eventually become extinct, and the time it takes for this to occur increases for systems nearer the bifurcation. When noise is incorporated, a stochastic amplitude equation for the even mode can be derived, which can be analyzed to reveal bump extinction time both below and above the saddle-node.

\end{abstract}

\section{Introduction}

Continuum neural fields are a well-accepted model of spatiotemporal neuronal activity evolving within in vitro and in vivo brain tissue \cite{coombes05,bressloff12}. Wilson and Cowan initially introduced these nonlocal integrodifferential equations to model activity of neuronal populations in terms of mean firing rates \cite{wilson72}. While they discount the intricate dynamics of neuronal spiking, these models can qualitatively capture a wide range of phenomena such as propagating activity waves observed in disinhibited slice preparations \cite{richardson05,pinto05,huang04,pinto01}. Neural field models exhibit a wide variety of spatiotemporal dynamics including traveling waves, Turing patterns, stationary pulses, breathers, and spiral waves \cite{ermentrout98,folias04,laing05,hutt10,coombes12b}. A distinct advantage of utilizing these continuum equations to model large-scale neural activity is that many analytical methods for studying their behavior can be adapted from nonlinear partial differential equations (PDEs) \cite{bressloff12}. Recently, several authors have explored the impact of stochasticity on spatiotemporal patterns in neural fields \cite{hutt08,bressloff12b,kilpatrick13} by employing techniques originally used to study stochastic front propagation in reaction-diffusion systems \cite{sagues07}. Typically, the approach is to perturb about a linearly stable solution of the deterministic system, under the assumption of weak noise. However, some recent efforts have been aimed at understanding the impact of noise on patterns near bifurcations \cite{hutt08,kilpatrick14}.

In this work, we are particularly interested in how noise interacts with stationary pulse (bump) solutions near a saddle-node bifurcation at which a branch of stable bumps and a branch of unstable bumps annihilate \cite{amari77}. Bumps are commonly utilized as a model of persistent and tuned neural activity underlying spatial working memory \cite{funahashi89,wimmer14}. This activity tends to last for a few seconds, after which it is extinguished, to allow for subsequent memories to be formed \cite{goldmanrakic95}. One possible way to terminate these sustained activity patterns is by transiently synchronizing the spiking patterns of excitatory neurons that participate in the signal \cite{gutkin01}. Another proposed mechanism for terminating persistent activity is a strong and brief global inhibitory signal, which would drive the system from the stable bump state to a stable uniform quiescent state \cite{compte00}. In terms of neural field and spiking models, this can be thought of as momentarily raising the firing threshold of the system, temporarily driving it beyond the saddle-node bifurcation from which the stable bump emerges. 

We focus on a scalar neural field model that supports stationary bump solutions for appropriate choices of parameters and constituent functions \cite{amari77,coombes05}:
\begin{align}
\frac{\pd u(x,t)}{\pd t} &= - u(x,t) + \int_{\Omega} w(x-y) f(u(y,t)) \d y \label{bfield}
\end{align}
where $u(x,t)$ is the total synaptic input arriving to location $x$ and time $t$, and $w(x-y)$ describes the strength (amplitude) and polarity (sign) of synaptic connections from neurons at location $y$ to neurons at location $x$. We assume $w(x)$ is an even-symmetric function $w(x) = w(-x)$ with a bounded integral $\int_{\Omega} w(x) \d x$ over the spatial domain $x \in \Omega = (-x_{\infty}, x_{\infty})$. The nonlinearity $f(u)$ is a firing rate function, which we take to be the sigmoid \cite{wilson72}
\begin{align}
f(u) = \frac{1}{1 + \e^{- \eta (u - \theta)}}, \label{sig}
\end{align}
and we also find it useful to take the high gain limit $\eta \to \infty$, in which case:
\begin{align}
f(u) = H(u - \theta ) =  \left\{ \begin{array}{cl} 1 & : u \geq \theta, \\ 0 & : u< \theta, \end{array} \right.  \label{H}
\end{align}
allowing for analytical tractability in several of our calculations. It is important to note that (\ref{bfield}) neglects several known features of neuronal networks including spike rate adaptation \cite{hansel98}, propagation delays \cite{hutt03}, synaptic depression \cite{kilpatrick10}, and refractoriness \cite{curtu01}. Thus, we assume we are focusing on a network where these effects are weak enough as to not impact our main results.

Amari was the first to analyze (\ref{bfield}) in detail, showing that when $f(u)$ is defined to be a Heaviside function (\ref{H}), the network supports stable stationary bump solutions when the weight function $w(x)$ is a lateral inhibitory (Mexican hat) distribution satisfying: (i) $w(x) > 0$ for $x \in [0,x_0)$ with $w(x_0) = 0$; (ii) $w(x)<0$ for $x \in (x_0, x_{\infty})$; (iii) $w(x)$ is decreasing on $[0,x_0]$; and (iv) $w(x)$ has a unique minimum on $[0,x_{\infty})$ at $x = x_1$ with $x_1>x_0$ and $w(x)$ strictly increasing on $(x_1,x_{\infty})$ \cite{amari77}. Based on restrictions (i)-(iv), Amari made use of the integral of the weight function
\begin{align}
W(x) \equiv \int_0^{x} w(y) \d y  \label{Wint}
\end{align}
to prove some of the main results of his seminal work. For instance, it is clear that $W(0) = 0$ and $W(x) = -W(-x)$ based on the above assumptions. Moreover, there will be a single maximum of the function $W(x)$ on the interval $(0,x_{\infty})$ given at $x = x_0$, i.e. $W_{max} = {\rm max}_x W(x)  = W(x_0)$, due to conditions (i) and (ii), and $w(x_0) = 0$. When $\theta < W(x_0)$ there are two bump solutions: one stable and one unstable (up to translation symmetry), and when $\theta > W(x_0)$ there are no bump solutions to (\ref{bfield}). When $\theta = \theta_c \equiv W(x_0)$, there is a single marginally stable bump solution. It is at this point that the two branches (stable and unstable) of bump solutions meet and annihilate in a {\em saddle-node bifurcation} (Fig. \ref{fig1}). Dynamics of (\ref{bfield}) for values of $\theta$ beyond this saddle-node bifurcation evolve to quasi-stationary solutions resembling the {\em ghost} of the bump at $\theta_c$, lasting for a period of time inversely related to $\sqrt{|\theta - \theta_c|}$ \cite{strogatzbook}. A principled exploration of these dynamics (section \ref{detsys}) is one of the primary goals of this paper.

As mentioned, the neural field equation (\ref{bfield}) in the absence of noise has been analyzed extensively \cite{amari77,ermentrout98,coombes05}. We expand upon these previous studies by also exploring the impact of noise on stationary bump solutions to (\ref{bfield}) near a saddle-node bifurcations (section \ref{stochsys}). Additive noise is incorporated, so that the evolution of the neural field is now described by the spatially extended Langevin equation \cite{laing01b,brackley07,hutt08,bressloff12}:
\begin{align}
\d u(x,t) = \left[ - u(x,t) + \int_{\Omega} w( x - y) f(u(y,t)) \d y \right] \d t + \epsilon \d W(x,t),  \label{nfield}
\end{align}
where the term $\d W(x,t)$ is the increment of a spatially varying Wiener process with mean defined by $\langle \d W(x,t) \rangle = 0$ and correlations $\d W(x,t) \d W(y,s) \rangle = C(x-y) \delta (t-s) \d t \d s$ and $\epsilon$ describes the amplitude of the noise, assumed to be weak ($\epsilon \ll 1$). The function $C(x-y)$ describes the spatial correlation in each noise increment between two points $x,y \in \Omega$. 

\section{Slow bump extinction in the deterministic system}
\label{detsys}

We begin by examining the dynamics of stationary bump solutions near a saddle-node bifurcation, where a stable and unstable branch of solutions annihilate. Our initial analysis focuses on the noise-free case $W(x,t) \equiv 0$, allowing us to derive an amplitude equation that approximates the evolution of the bump height. Linearization of bumps in (\ref{bfield}) typically reveals that they are marginally stable to translating perturbations, so the overall stability is characterized by the stability to even perturbations that expand/contract the bump \cite{ermentrout98}. Our analysis will emphasize the region of parameter space near where bumps are marginally stable to even perturbations.

\subsection{Existence and stability of bumps}

\begin{figure}
\begin{center} \includegraphics[width=6cm]{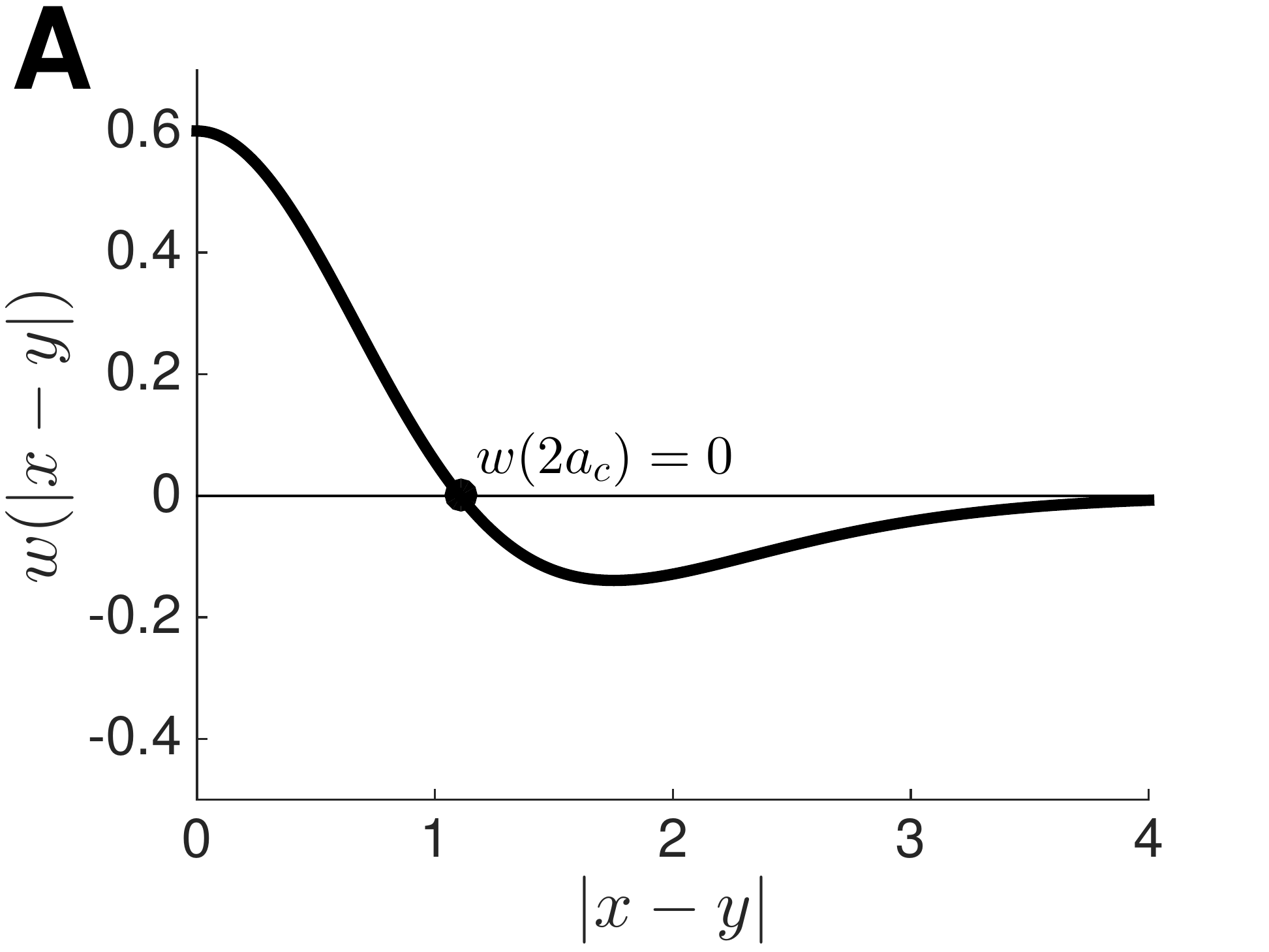} \includegraphics[width=6cm]{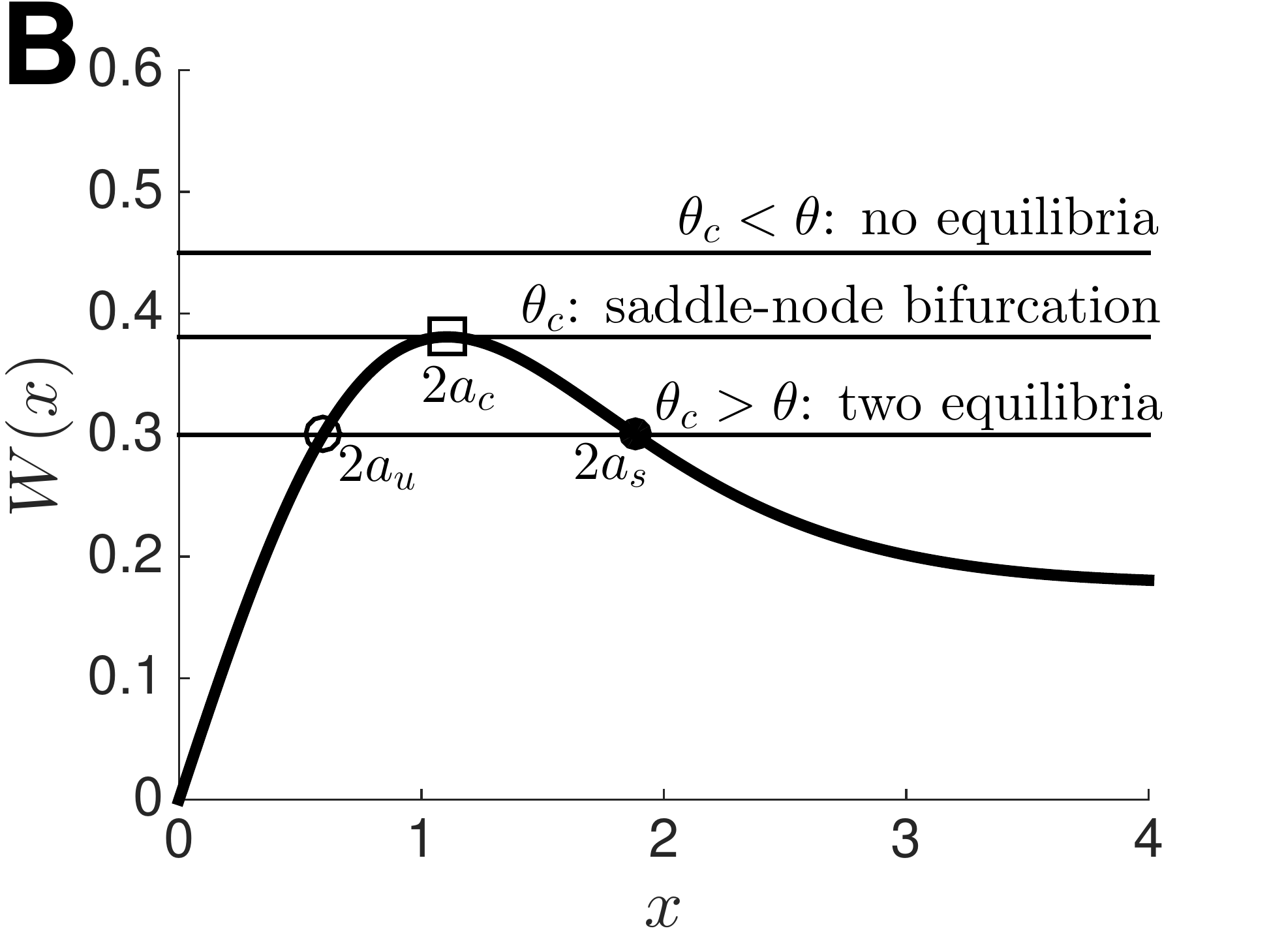} \end{center}
\caption{Saddle-node bifurcation of bumps in (\ref{bfield}) with a Heaviside firing rate function (\ref{H}). ({\bf A}) Difference of Gaussians weight function $w(x) = \e^{-x^2} - A \e^{-x^2/\sigma^2}$ has a Mexican hat profile with $A = 0.4<1$ and $\sigma = 2 > 1$. The critical bump half-width $a_c$ at the saddle-node satisfies the relation $w(2a_c) = 0$. ({\bf B}) The weight function integral (\ref{Wint}) determines the bump half-widths $a$. When $\theta$ is below the critical threshold $\theta_c$ at the saddle-node, there are two stationary bump solutions to (\ref{bfield}): one stable $a_s$ and one unstable $a_u$. When $\theta > \theta_c$, there are zero equilibria, but the dynamics of (\ref{bfield}) are slow in the bottleneck near $U_c(x)$.}
\label{fig1}
\end{figure}

We now briefly review existence and stability results for stationary bump solutions to the neural field equation (\ref{bfield}). These results are analogous to those presented in \cite{amari77,veltz10,kilpatrick13}. For transparency, we focus on the case of a Heaviside firing rate function (\ref{H}). This allows us to cast bump stability in terms of a finite dimensional set of equations, focusing on the evolution of the two edge interfaces of the bump \cite{amari77,coombes12}. Assuming a stationary solution $u(x,t) = U(x)$, we find (\ref{bfield}) requires
\begin{align}
U(x) = \int_{\Omega} w(x-y)H(U(y) - \theta) \d y.  \label{bmpeqn}
\end{align}
Given a unimodal bump solution $U(x)$, without loss of generality, we can fix the center and peak of the bump to be at the origin $x=0$. In the case of even-symmetric bumps $U(x) = U(-x)$ \cite{amari77}, we will have the conditions for the bump half-width $a$: $U(x) > \theta$ for $x \in (-a,a)$, $U(x)< \theta$ for $x \in \Omega \backslash [-a,a]$, and $U( \pm a) = \theta$. In this case, (\ref{bmpeqn}) becomes
\begin{align*}
U(x) = \int_{-a}^a w(x-y) \d y = \int_{x-a}^{x+a} w(y) \d y = \int_0^{x+a} w(y) \d y - \int_0^{x-a} w(y) \d y.
\end{align*}
By utilizing the integral function (\ref{Wint}), we can write the even-symmetric solution
\begin{align}
U(x) = W(x+a) - W(x-a).  \label{bumpW}
\end{align}
To determine the half-width $a$, we require the threshold conditions $U(\pm a) = \theta$ of the solution (\ref{bumpW}) to yield
\begin{align*}
U(a) = W(2a) = \int_0^{2a} w(y) \d y = \theta.
\end{align*}
Note that when $\theta < W_{max} = {\rm max}_x W(x)$, there will be a stable and unstable bump solution to (\ref{bfield}). When $\theta = \theta_c \equiv W_{max}$, there is a single marginally stable bump solution $U_c(x)$ to (\ref{bfield}), as illustrated in Fig. \ref{fig1}{\bf B}. Differentiating $W(2a)$ by its argument yields $W'(2a_c) = w(2a_c) \equiv 0$ as an implicit equation for the half-width $a_c$ at this criticality. Utilizing the notation of Amari condition (i), we have that $a_c = x_0/2$. Note, the relation $w(2a_c) = 0$ is explicitly solvable for $a_c$ for several typical lateral inhibitory type weight functions. For instance, in the case of the difference of Gaussians $w(x) = \e^{-x^2} - A \e^{-x^2/\sigma^2}$ on $x \in (-\infty, \infty)$ \cite{amari77}, we have $a_c = \sigma \sqrt{\ln (1/A)}/ \left[ 2 \sqrt{\sigma^2 - 1} \right]$ and $\theta_c = \frac{\sqrt{\pi}}{2} \left[ \erf (2a_c) - A \sigma \erf (2a_c/\sigma) \right]$. For the ``wizard hat" $w(x) = (1-|x|)\e^{-|x|}$ on $x \in (-\infty,\infty)$ \cite{coombes05b}, we have $a_c = 1/2$ and $\theta_c = \e^{-1}$. For a cosine weight $w(x) = \cos (x)$ on the periodic domain $x \in [- \pi, \pi]$ \cite{kilpatrick13}, we have $a_c = \pi /4$ and $\theta_c = 1$.

To characterize the stability of bump solutions to (\ref{bfield}), we will study the evolution of small smooth perturbations $\ve \bar{\psi} (x,t)$ ($\ve \ll 1$) to stationary bumps $U(x)$ by utilizing the Taylor expansion $u(x,t) = U(x) + \ve \bar{\psi} (x,t) + {\mc O}(\ve^2)$. By plugging this expansion into (\ref{bfield}) and truncating to ${\mc O}(\ve)$, we can derive an equation whose solutions constitute the family of eigenfunctions associated with the linearization of (\ref{bfield}) about the bump solution $U(x)$. We begin by truncating (\ref{bfield}) to ${\mc O}(\ve )$ assuming $u$ is given by the above expansion and that the nonlinearity $f(u)$ is given by the Heaviside function (\ref{H}), so
\begin{align}
\frac{\pd \bar{\psi}(x,t)}{\pd t} = - \bar{\psi}(x,t) + \int_{\Omega} w(x-y) H'(U(y) - \theta) \bar{\psi}(y,t) \d y, \label{dpsi1}
\end{align}
and we can differentiate the Heaviside function, in the sense of distributions, by noting $H(U(x) - \theta) = H(x+a) - H(x-a)$, so
\begin{align*}
\delta (x+a) - \delta (x-a) = \frac{\d H(U(x) - \theta)}{\d x} & = H'(U(x) - \theta)U'(x),
\end{align*}
which we can rearrange to find
\begin{align}
H'(U(x) - \theta) &= \frac{\delta (x+a) - \delta (x-a)}{U'(x)} = \frac{1}{|U'(a)|} \left( \delta (x+a) + \delta (x-a) \right).  \label{Hdiff}
\end{align}
Upon applying the identity (\ref{Hdiff}) to (\ref{dpsi1}), we have
\begin{align}
\frac{\pd \bar{\psi}(x,t)}{\pd t} = - \bar{\psi}(x,t) + \gamma \left[ w(x+a) \bar{\psi}(-a,t) + w(x-a) \bar{\psi}(a,t) \right], \label{dpsi2}
\end{align}
where $\gamma^{-1} = |U'(a)| = w(0) - w(2a)$. One class of solutions, such that $\bar{\psi}(\pm a,t) = \bar{\psi}(\pm a,0) = 0$, lies in the essential spectrum of the linear operator that defines (\ref{dpsi2}). In this case, $\bar{\psi}(x,t) = \bar{\psi}(x,0) \e^{-t}$, so perturbations of this type do not contribute to any instabilities of the stationary bump $U(x)$ \cite{guo05}. Assuming separable solutions $\bar{\psi}(x,t) = b(t) \psi (x)$, we can characterize the remaining solutions to (\ref{dpsi2}). In this case, $b'(t) = \lambda b(t)$, so $b(t) = \e^{\lambda t}$ where $\lambda \in \R$, and
\begin{align}
(\lambda + 1) \psi (x) = \gamma \left[ w(x+a) \psi(-a) + w(x-a) \psi (a) \right]. \label{lampsi}
\end{align}
Solutions to (\ref{lampsi}) that do not satisfy the condition $\psi (\pm a) \equiv 0$ can be separated into two classes: (i) odd $\psi (a) = - \psi (-a)$ and (ii) even $\psi(a) = \psi (-a)$. This is due to the fact that the equation (\ref{lampsi}) implies the function $\psi (x)$ is fully specified by its values at $x= \pm a$. Thus, we need only concern ourselves with these two points, yielding the two-dimensional linear system
\begin{subequations}
\begin{align} \label{psieigf}
( \lambda + 1) \psi (-a) & = \gamma \left[ w(0) \psi (-a) + w(2a) \psi (a) \right] \\
( \lambda + 1) \psi (a) & = \gamma \left[ w(2a) \psi (-a) + w(0) \psi (a) \right].
\end{align}
\end{subequations}
For odd solutions $\psi (a) = - \psi (-a)$, the eigenvalue
\begin{align*}
\lambda_o = -1 + \gamma \left[ w(0) - w(2a) \right] = -1 + \frac{w(0) - w(2a)}{w(0) - w(2a)} = 0,
\end{align*}
reflecting the fact that (\ref{bfield}) is translationally symmetric, so bumps are marginally stable to perturbations that translate their position. Even solutions $\psi (a) = \psi (-a)$ have associated eigenvalue
\begin{align*}
\lambda_e = -1 + \gamma \left[ w(0) + w(2a) \right] = -1 + \frac{w(0) +w(2a)}{w(0) - w(2a)} = \frac{2 w(2a)}{w(0) - w(2a)}.
\end{align*}
Thus, when $\theta < \theta_c$, the wide bump $a_s > a_c$ will be linearly stable to expanding/contracting perturbations since $w(2a_s)<0$ due to Amari's condition (ii) \cite{amari77}. The narrow bump $a_u< a_c$ is linearly unstable to such perturbations since $w(2a_u)>0$ due to condition (i). When $\theta = \theta_c$, we have $w(2a_c) = 0$ so that $\lambda_e = 0$ and $|U'(\pm a_c)| = w(0)$.

In anticipation of our derivations of amplitude equations, we define the eigenfunctions at the criticality $\theta = \theta_c$. Utilizing the fact that $|U'(\pm a_c)| = w(0)$ and the linear system (\ref{psieigf}), we have that the odd eigenfunction at the bifurcation is
\begin{align}
\psi_o(x) &= \frac{1}{w(0)} \left[ w(x-a_c) - w(x+a_c) \right],  \label{oddpsi}
\end{align}
and the even eigenfunction is
\begin{align}
\psi_e(x) &= \frac{1}{w(0)} \left[ w(x-a_c) + w(x+a_c) \right].  \label{evenpsi}
\end{align}
Note, this specifies that $\psi_e(\pm a) = \psi_o(a) = - \psi_0(-a) = 1$. Furthermore, we will find it useful to compute the derivatives
\begin{align*}
\psi_o'(x) &= \frac{1}{w(0)} \left[ w'(x-a) - w'(x+a) \right],
\end{align*}
which is even ($\psi_o'(-a_c) = \psi_o'(a_c)$), and
\begin{align*}
\psi_e'(x) &= \frac{1}{w(0)} \left[ w'(x-a) + w'(x+a) \right],
\end{align*}
which is odd ($\psi_e'(-a_c) = -\psi_e'(a_c)$). Lastly, we note that we will utilize the fact that, for even symmetric functions, $w'(0) = 0$, so $\psi_o'(\pm a_c) = \psi_e'(a_c) = - \psi_e'(-a_c) = w'(2a_c)/w(0)$. 

\subsection{Saddle-node bifurcation of bumps}

Motivated by the above linear stability analysis, we now carry out a nonlinear analysis in the vicinity of the saddle-node bifurcation from which the stable and unstable branches of stationary bumps emanate. Specifically, we will perform a perturbation expansion about the bump solution $U_c(x)$ at the critical threshold value $\theta_c$. We therefore define $\theta = \theta_c + \mu \ve^2$, $\ve \ll 1$, so that $\mu$ is a bifurcation parameter determining the distance of $\theta$ from the saddle-node bifurcation point. As demonstrated above, the linear stability problem for $U_c(x)$ reveals two zero eigenvalues $\lambda_o = \lambda_e = 0$ associated with the odd $\psi_o$ and even $\psi_e$ eigenfunctions (\ref{oddpsi}) and (\ref{evenpsi}), respectively. Our analysis employs the ansatz:
\begin{align}
u(x,t) = U_c(x) + \ve A_e(\tau) \psi_e(x) + \ve^2 A_o( t) \psi_o(x) +  \ve^2 u_2(x, \tau) + {\mc O}(\ve^3),  \label{banz}
\end{align}
where $\tau = \ve t$ is a temporal rescaling that reflects the vicinity of the system to a saddle-node bifurcation associated with the even expanding/contracting eigenmode $\psi_e$ \cite{strogatzbook}. Similar expansions have been utilized in the analysis of bifurcations for spatial patterns in reaction-diffusion systems \cite{schutz95,bode97} and neural field models \cite{bressloff04,hutt05,venkov07}.  Upon plugging (\ref{banz}) into (\ref{bfield}) and expanding in orders of $\ve$, we find that at ${\mc O}(1)$, we simply have the stationary bump equation (\ref{bmpeqn}) at $\theta = \theta_c$. Proceeding to ${\mc O}( \ve)$, we find
\begin{align*}
0 = A_e(\tau) \left[ \int_{\Omega} w(x-y) H'(U_c(y) - \theta_c) \psi_e(y) \d y - \psi_e(x) \right],
\end{align*}
so we can use (\ref{Hdiff}) to write
\begin{align}
0 = A_e(\tau) \left[ \frac{1}{w(0)}(w(x+a) \psi_e(-a) + w(x-a) \psi_e(a)) - \psi_e(x) \right].  \label{ordepssum}
\end{align}
The right hand side of (\ref{ordepssum}) vanishes due to the formula for the even (\ref{evenpsi}) eigenfunction associated with the stability of the bump $U_c(x)$. At ${\mc O}(\ve^2)$, we obtain an equation for higher order term $u_2$:
\begin{align}
{\mc L} \left[ A_o \psi_o + u_2 \right] = & A_e' \psi_e + A_o' \psi_o  +\mu \int_{\Omega} w(x-y) H'(U_c(y)-\theta_c) \d y \label{u2eqn} \\ 
& -  \frac{A_e^2}{2} \int_{\Omega} w(x-y) H''(U_c(y)-\theta_c) \psi_e (y)^2 \d y,  \nonumber
\end{align}
where ${\mc L}$ is the non-self-adjoint linear operator
\begin{align}
{\mc L} u (x) = - u(x) + \int_{\Omega} w(x-y) H'(U_c(y) - \theta_c) u(y) \d y.  \label{linop}
\end{align}
Both $\psi_o(x)$ and $\psi_e(x)$ lie in the nullspace ${\mc N}({\mc L})$, as demonstrated in the previous section by identifying solutions to (\ref{dpsi1}). Thus, the $\psi_o$ terms on the left hand side of (\ref{u2eqn}) vanish. We can ensure a bounded solution to (\ref{u2eqn}) exists by requiring that the right hand side be orthogonal to all elements of the nullspace of the adjoint operator ${\mc L}^*$. The adjoint is defined with respect to the $L^2$ inner product
\begin{align}
\langle {\mc L} u, v \rangle = \int_{\Omega} \left[ {\mc L} u (x) \right] v(x) \d x = \int_{\Omega} u(x) \left[ {\mc L}^* v(x) \right] \d x = \langle u, {\mc L}^* v \rangle.  \label{l2ip}
\end{align}
Thus, we find
\begin{align}
{\mc L}^* v(x) = - v(x) + H'(U_c(x) - \theta_c) \int_{\Omega} w(x- y) v(y) \d y,  \label{adjop}
\end{align}
defined in the sense of distributions under the $L^2$ inner product given in (\ref{l2ip}). It is straightforward to show that $\varphi_o : = H'(U_c- \theta_c) \psi_o$ and $\varphi_e : = H'(U_c- \theta_c) \psi_e$ lie in the nullspace of ${\mc L}^*$. Components of ${\mc N}({\mc L}^*)$ are defined by the equation
\begin{align}
v(x) = H'(U_c(x) - \theta_c) \int_{\Omega} w(x-y) v(y) \d y.   \label{Lsns}
\end{align}
To show $\varphi_o, \varphi_e \in {\mc N}({\mc L}^*)$, we simply plug these formulas into (\ref{Lsns}) to find
\begin{align*}
H'(U_c(x)- \theta_c) \psi_j(x) = H'(U_c(x) - \theta_c) \int_{\Omega} w(x-y) H'(U_c(y)- \theta_c) \psi_j (y) \d y,
\end{align*}
for $j=o,e$, which is true due to the fact that $\psi_o$ and $\psi_e$ lie in ${\mc N}({\mc L})$. Thus, we will impose solvability of (\ref{u2eqn}) by taking the inner product of both sides of the equation with respect to $\varphi_o : = H'(U_c- \theta_c) \psi_o$ and $\varphi_e : = H'(U_c- \theta_c) \psi_e$ yielding
\begin{align}
0 = & \left\langle \varphi_j, A_e' \psi_e + A_o' \psi_o + \mu w*H'(U_c-\theta_c) -  \frac{A_e^2}{2}w*\left[ H''(U_c-\theta_c) \psi_e^2 \right] \right\rangle, \label{psisolv1}
\end{align}
for $j=o,e$, where we have defined the convolution $w*F = \int_{\Omega} w(x-y) F(y) \d y$. Due to odd-symmetry, terms of the form $\langle H'(U_c- \theta_c) \psi_j, \psi_k \rangle$, $j \neq k$, vanish. In a similar way, the term $\langle H'(U_c- \theta_c) \psi_o, w*H'(U_c-\theta_c) \rangle$ vanishes due to odd-symmetry. Isolating the temporal derivatives $A_j'$ in (\ref{psisolv1}), we find that the amplitudes $A_j$ ($j=o,e$) satisfy the following fast-slow system of nonlinear differential equations
\begin{subequations} \label{snamp}
\begin{align} 
\frac{\d A_o}{\d t} &=   \frac{\left\langle \varphi_o , w*\left[H''(U_c-\theta_c)\psi_e^2 \right] \right\rangle}{2 \langle \varphi_o, \psi_o \rangle} A_e(\tau)^2, \\
\frac{\d A_e}{\d \tau} &= - \mu \frac{\langle \varphi_e, w*H'(U_c-\theta_c) \rangle}{\langle \varphi_e, \psi_e \rangle} + \frac{\left\langle \varphi_e , w*\left[H''(U_c-\theta_c) \psi_e^2 \right] \right\rangle}{2 \langle \varphi_e, \psi_e \rangle} A_e(\tau)^2.
\end{align}
\end{subequations}

With the system (\ref{snamp}) in hand, we can determine the long term dynamics of the amplitudes as the bifurcation parameter $\mu$ is varied. We begin by computing the constituent components of the right hand sides, using properties of the eigenfunctions $\psi_o$ and $\psi_e$. To start, we will compute the second derivative $H''(U_c - \theta_c)$, which appears in the coefficient of the quadratic term $A_e^2$. Differentiating the function $H(U_c(x) - \theta_c)$ twice with respect to $x$, using the chain and product rule, we find the following formula
\begin{align*}
\frac{\d^2 H(U_c(x) - \theta_c)}{\d x^2} &= (U_c'(x))^2 H''(U_c(x) - \theta_c)  + U_c''(x) H'(U_c(x) - \theta_c)  \\
&= (U_c'(x))^2 H''(U_c(x) - \theta_c)  +  \frac{U_c''(x)}{U_c'(x)} \frac{\d H(U_c(x) - \theta_c)}{\d x},
\end{align*} 
where we have applied the identity (\ref{Hdiff}) for the first derivative $H'(U- \theta)$. Rearranging terms, we find that
\begin{align}
H''(U_c - \theta_c) = \frac{1}{U_c'(x)^2} \frac{\d^2 H(U_c(x) - \theta_c)}{\d x^2} - \frac{U_c''(x)}{|U_c'(a)|^3} \left[ \delta (x+a_c) + \delta (x-a_c) \right].  \label{H2diff1}
\end{align}
We can further specify the formula (\ref{H2diff1}) by differentiating $\frac{\d H(U_c - \theta_c)}{\d x} = \delta (x+a_c) - \delta (x-a_c)$ with respect to $x$ to yield
\begin{align*}
\frac{\d^2 H(U_c - \theta_c)}{\d x^2} = \delta' (x+a_c) - \delta' (x-a_c), 
\end{align*}
where $\delta'(x - x_0)$ is defined, in the sense of distributions, for any smooth function $F(x)$ by using integration-by-parts \cite{keenerbook}:
\begin{align*}
\int_{\Omega} \delta'(x-x_0) F(x) \d x = - \int_{\Omega} \delta(x-x_0) F'(x) \d x = -F'(x_0).
\end{align*}
Furthermore, we note that the spatial derivatives $|U_c'( \pm a_c)| = w(0)$ and $U_c''(x) = w'(x+a_c) - w'(x-a_c)$. Even symmetry of $w(x)$ mandates that $w'(x) = -w'(-x)$ and $w'(0) = 0$, so $U_c''(\pm a_c) = w'(2a_c)$. Thus, we can at last write
\begin{align}
H''(U_c - \theta_c) = \frac{\delta'(x+a_c) - \delta' (x-a_c)}{w(0)^2} - \frac{w'(2a_c)\left[ \delta (x+a_c) + \delta (x-a_c) \right]}{w(0)^3}.  \label{H2difid}
\end{align}
Computing the inner products in (\ref{snamp}) then simply amounts to evaluating the integrals in the sense of distributions. First, we use (\ref{Hdiff}) to note
\begin{align*}
\langle \varphi_j, \psi_j \rangle = \int_{\Omega} \psi_j(x)^2 H'(U_c(x)-\theta_c) \d x = \gamma \left[ \psi_j(a_c)^2 + \psi_j(-a_c)^2 \right] = \frac{2}{w(0)},
\end{align*}
for $j=o,e$, since $\psi_e(\pm a_c) = \psi_o(a_c) = - \psi_o(-a_c)=1$. Furthermore,
\begin{align}
\langle \varphi_e, w*H'(U_c-\theta_c) \rangle &= \int_{\Omega} \int_{\Omega} w(x-y) \psi_e(x) H'(U_c(x) - \theta_c) H'(U_c(y)-\theta_c) \d y \d x \nonumber \\
& = \gamma^2 \int_{\Omega} \left[ \sum_{a=\pm a_c} w(x+a) \right] \psi_e (x) \left[ \sum_{a=\pm a_c} \delta(x+a) \right] \d x \nonumber \\
&= \gamma^2 \left[ \psi_e(a_c) + \psi_e(-a_c) \right] \cdot \left[ w(0) + w(2a_c) \right] = \frac{2}{w(0)},  \label{bifparip}
\end{align}
where we have utilized $\psi_e(\pm a_c) = 1$ and $w(2a_c) \equiv 0$. Finally, we compute the quadratic terms using the identity (\ref{H2difid}), starting with
\begin{align}
\langle \varphi_o, w* \left[ H''(U_c - \theta_c) \psi_e^2 \right] \rangle &= \int_{\Omega} \int_{\Omega} w(x-y) \varphi_o(x) H''(U_c(y) - \theta_c) \psi_e(y)^2 \d y \d x \nonumber \\
&= \gamma \sum_{a=\pm a_c} \psi_o(a) \int_{\Omega} w(a - y) H''(U_c(y)-\theta_c) \psi_e(y)^2 \d y, \label{qoddc1}
\end{align}
and we note that individual terms under the integral from the sum defining (\ref{H2difid}) are
\begin{align*}
\int_{\Omega} w(-a_c - y) \delta'(y+a_c) \psi_e(y)^2 \d y & = w'(0) \psi_e(-a_c)^2 - 2 w(0) \psi_e'(-a_c) \psi_e(-a_c) \\
& = 2w'(2a_c),  \\
\int_{\Omega} w(a_c - y) \delta'(y+a_c) \psi_e(y)^2 \d y &= w'(2a_c) \psi_e(-a_c)^2 - 2 w(2 a_c) \psi_e'(-a_c) \psi_e(-a_c) \\
&= w'(2a_c), \\
\int_{\Omega} w(-a_c - y) \delta'(y-a_c) \psi_e(y)^2 \d y &= w'(-2a_c) \psi_e(a_c)^2 - 2 w(2a_c) \psi_e'(a_c) \psi_e(a_c) \\
& = - w'(2a_c), \\
\int_{\Omega} w(a_c - y) \delta'(y-a_c) \psi_e(y)^2 \d y &= w'(0) \psi_e(a_c)^2 - 2 w(0) \psi_e'(a_c) \psi_e(a_c) \\
& = - 2w'(2a_c), 
\end{align*}
for the terms involving the distributional derivative $\delta'(x-x_0)$, whereas the terms involving $\delta (x-x_0)$ are
\begin{align*}
\int_{\Omega} w(-a_c - y) \delta(y+a_c) \psi_e(y)^2 \d y & = w(0) \psi_e(-a_c)^2 = w(0), \\
\int_{\Omega} w(a_c - y) \delta (y+a_c) \psi_e(y)^2 \d y &= w(2a_c) \psi_e(-a_c)^2 = 0, \\
\int_{\Omega} w(-a_c - y) \delta(y-a_c) \psi_e(y)^2 \d y & = w(2a_c) \psi_e(a_c)^2 = 0, \\
\int_{\Omega} w(a_c - y) \delta(y-a_c) \psi_e(y)^2 \d y & = w(0) \psi_e(a_c)^2 = w(0).
\end{align*}
Thus, each integral term
\begin{align}
\int_{\Omega} w(-a_c-y) H''(U_c(y) - \theta_c) \psi_e(y)^2 \d y =& \frac{2 w'(2a_c)}{w(0)^2} \label{wmaH2} \\
\int_{\Omega} w(a_c-y) H''(U_c(y) - \theta_c) \psi_e(y)^2 \d y =&  \frac{2 w'(2a_c)}{w(0)^2}. \label{wpaH2}
\end{align}
Finally, using the fact that $\psi_o(a) = - \psi_o(-a) = 1$, we find that the two terms in the sum of (\ref{qoddc1}) cancel and the integral vanishes. Thus, $\langle \varphi_o, w*\left[ H''(U_c - \theta_c) \psi_e^2 \right] \rangle = 0$, so $A_o(t) \equiv \bar{A}_o$ is constant. On the other hand, computing the quadratic coefficient in the equation for $A_e$, we have
\begin{align}
\langle \varphi_e, w*\left[ H''(U_c - \theta_c) \psi_e^2 \right] \rangle &= \int_{\Omega} \int_{\Omega} w(x-y) \varphi_e(x)H''(U_c(y) - \theta_c) \psi_e(y)^2 \d y \d x \nonumber \\
&= \gamma \sum_{a = \pm a_c} \psi_e(a) \int_{\Omega} w(a-y) H''(U_c(y) - \theta_c) \psi_e(y)^2 \d y. \label{qevenc1}
\end{align} 
The integrals in (\ref{qevenc1}) are identical to those in (\ref{qoddc1}), so it is straightforward to compute, using (\ref{wmaH2}) and (\ref{wpaH2}) that
\begin{align*}
\langle \varphi_e, w*\left[ H''(U_c - \theta_c) \psi_e^2 \right] \rangle &= \gamma \left[ \frac{2 w'(2a_c)}{w(0)^2} +  \frac{2 w'(2a_c)}{w(0)^2}  \right] = \frac{4 w'(2a_c)}{w(0)^3}.
\end{align*}
Thus, we can at last compute all the terms in (\ref{snamp}), specifying that
\begin{subequations} \label{snamp2}
\begin{align}
\frac{\d A_o}{\d t} & = 0,  \label{sna1} \\
\frac{\d A_e}{\d \tau} &= -\mu - \frac{|w'(2a_c)|}{w(0)^2} A_e(\tau)^2, \label{sna2}
\end{align}
\end{subequations}
where we have noted the fact that $w'(2a_c)<0$ due to Amari's conditions (iii) and (iv) on the weight function $w(x)$ \cite{amari77}.

Equation (\ref{sna1}) reflects the translational symmetry of the original neural field equation (\ref{bfield}), so bumps are neutrally stable to translating perturbations $\psi_o$ regardless of the bifurcation parameter $\mu$. On the other hand, as the bifurcation parameter $\mu$ is changed, the dynamics of the even eigenmode $\psi_e$ reflect the relative distance to the saddle-node bifurcation at which point bumps are marginally stable to expanding/contracting perturbations. When $\mu <0$, there are two fixed points of equation (\ref{sna2}) at $A_e = \pm w(0) \sqrt{|\mu / w'(2a_c)|}$, corresponding to the pair of emerging stationary bump solutions which are wider ($+$) and narrower ($-$) than the critical bump $U_c$. As expected, the wide bump is linearly stable since a linearization of (\ref{sna2}) yields $\lambda_+ = - \sqrt{|\mu \cdot w'(2a_c)|}/w(0) < 0$, and the narrow bump is linearly unstable since $\lambda_- = + \sqrt{|\mu \cdot w'(2a_c)|}/w(0) > 0$. Crossing through the subcritical saddle-node bifurcation, we find that for $\mu \equiv 0$, there is a single fixed point $A_e \equiv 0$, which is marginally stable, since $\lambda_0 = 0$.

\begin{figure}
\begin{center} \includegraphics[width=6cm]{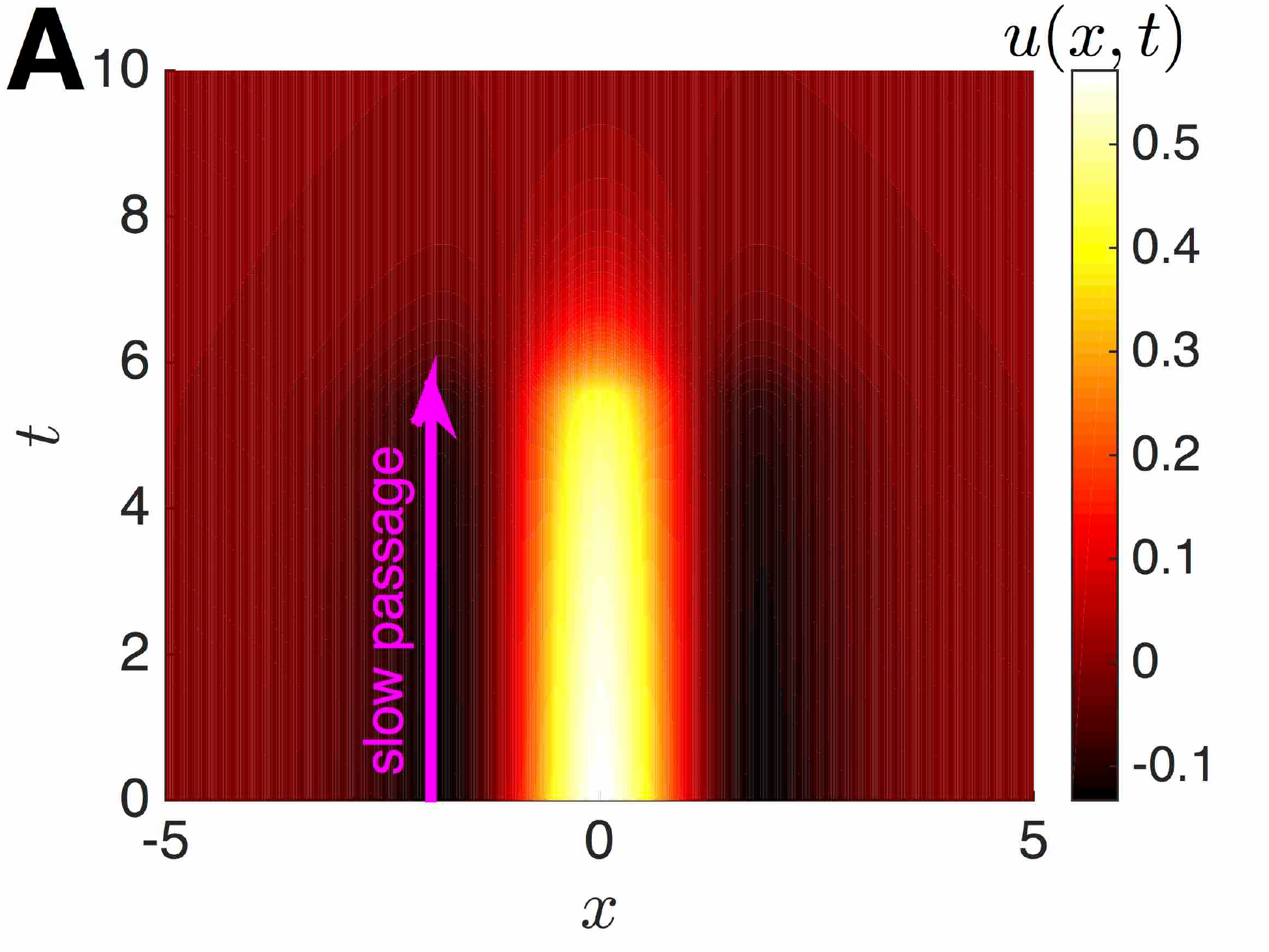} \includegraphics[width=6cm]{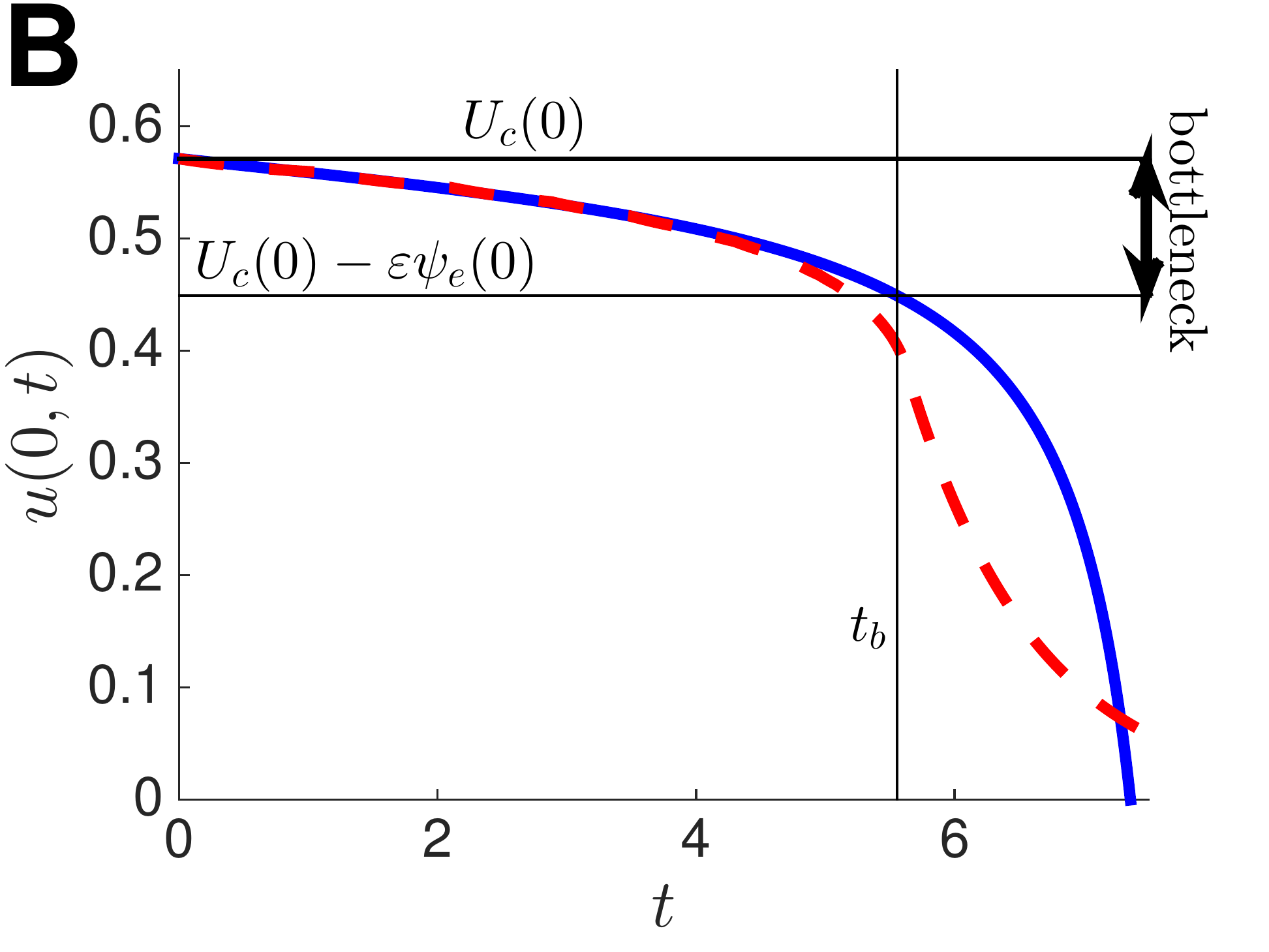}  \\ \includegraphics[width=6cm]{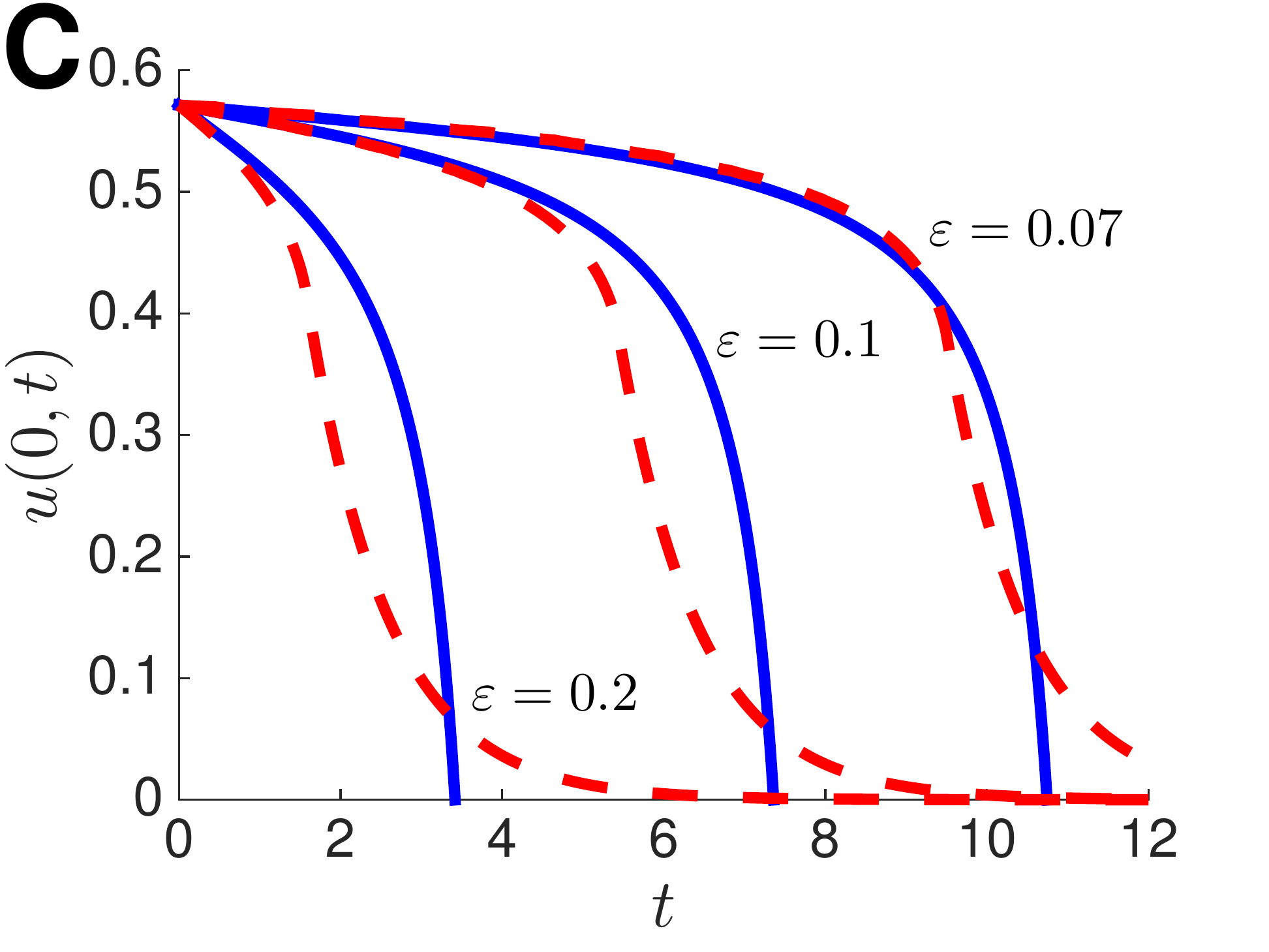}  \includegraphics[width=6cm]{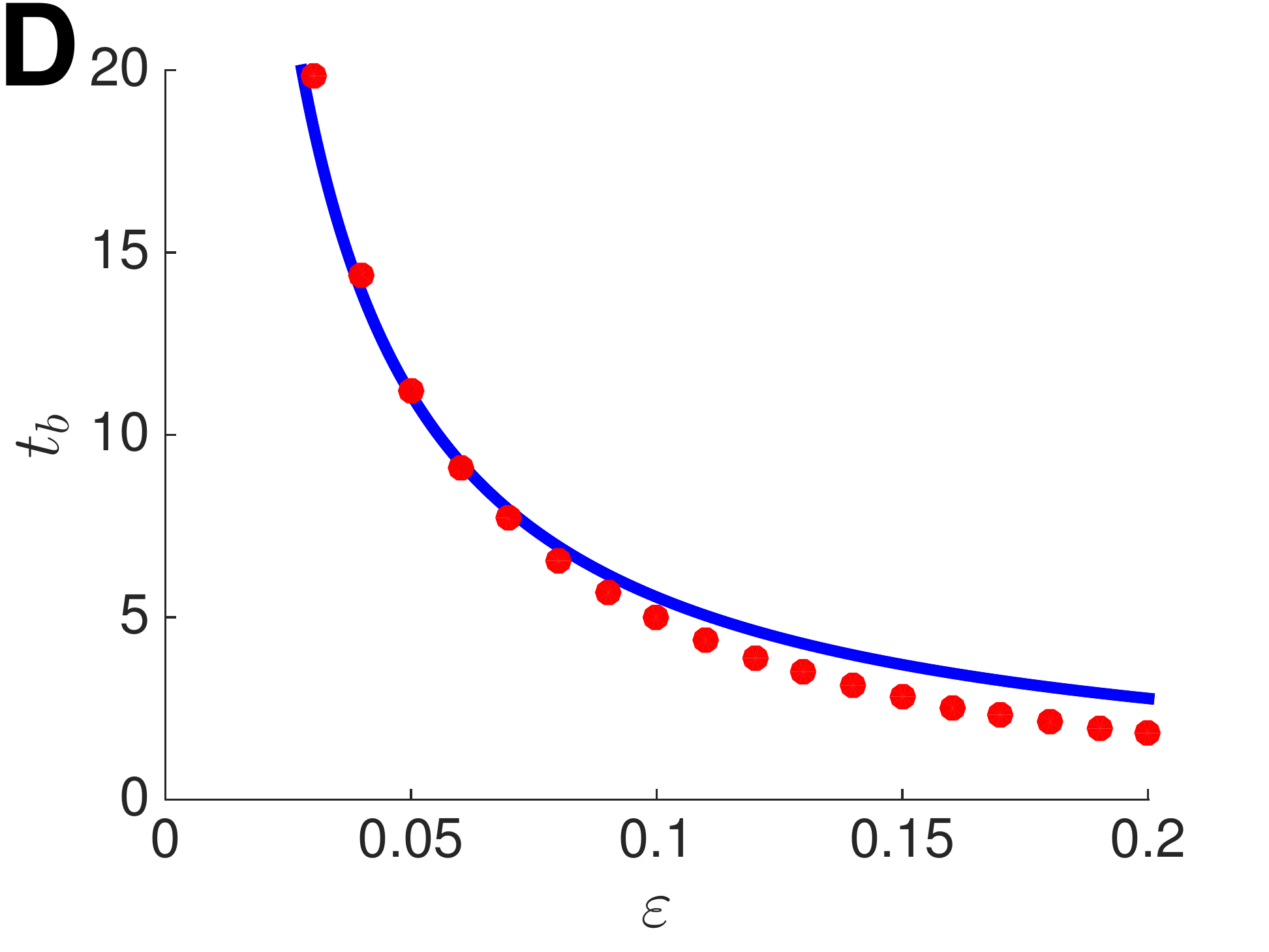} \end{center}
\caption{Slow passage of bumps on $x \in (-\infty, \infty)$ when $w(x) = \e^{-x^2} - A \e^{-x^2/\sigma^2}$. ({\bf A}) Slow passage of a transient bump by the {\em ghost} of the critical solution $U_c(x)$ when $\theta = \theta_c + \varepsilon^2$ for $\varepsilon = 0.1$ ($\mu =1$), $A=0.4$, and $\sigma = 2$. ({\bf B}) The peak of the bump $u(0,t)$ slowly decreases in amplitude until breaking down quickly in the vicinity of $A_e(t) = -1$. Note the theoretical formula for the amplitude (solid line) given by (\ref{Aetan}) matches the numerical simulation (dashed line) in the slow passage region. ({\bf C}) Amplitude of the even mode $A_e(t)$ slowly decreases with time. The duration of the bottleneck increases as the distance to the bifurcation is decreased by reducing $\varepsilon$. ({\bf D}) Comparison of the theory (solid) given by (\ref{btime}) to the numerically computed (dots) duration in the bottleneck (the crossing $A_e(t_b) = -1$).}
\label{fig2}
\end{figure}

\begin{figure}
\begin{center} \includegraphics[width=6cm]{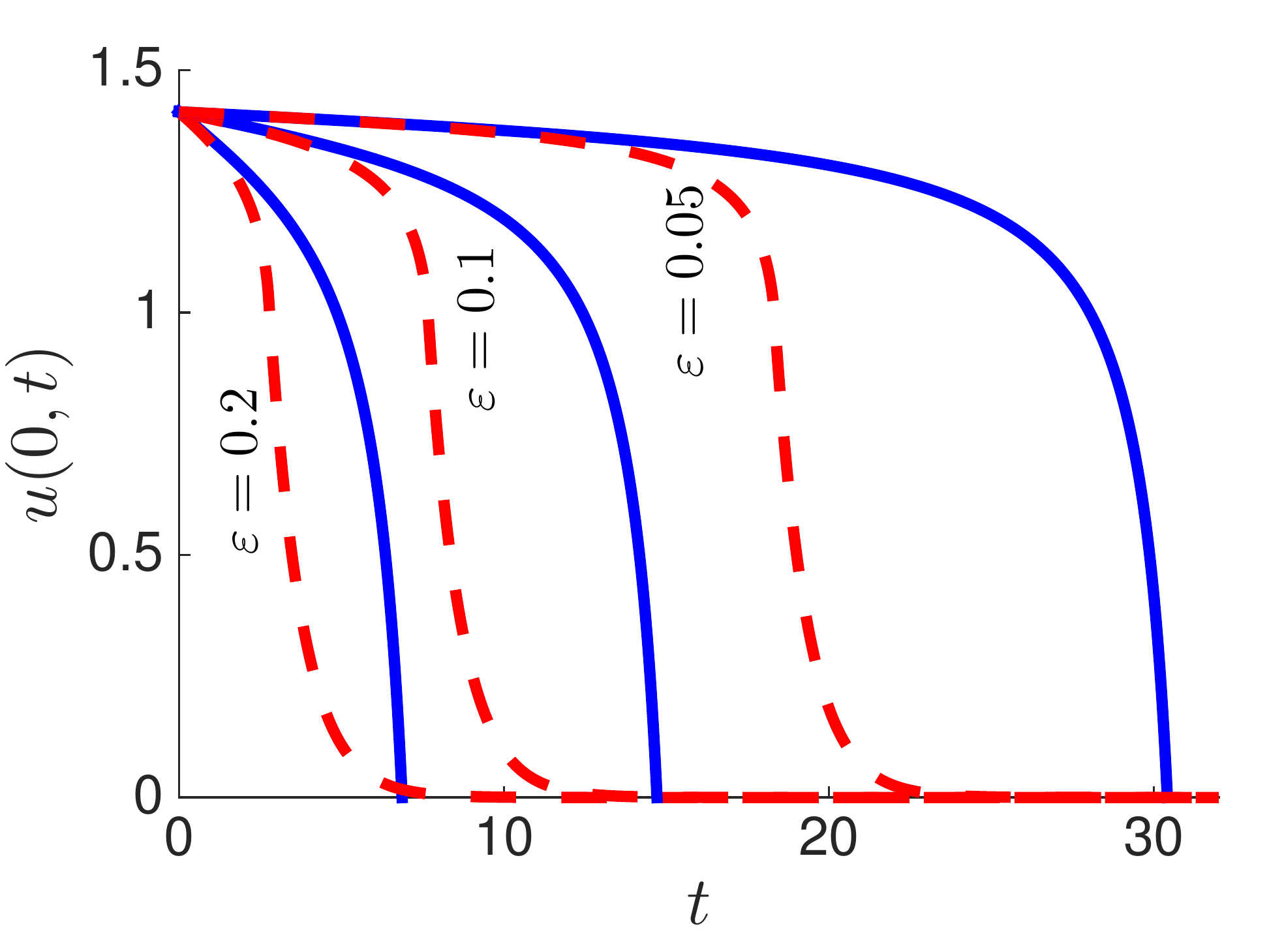} \includegraphics[width=6cm]{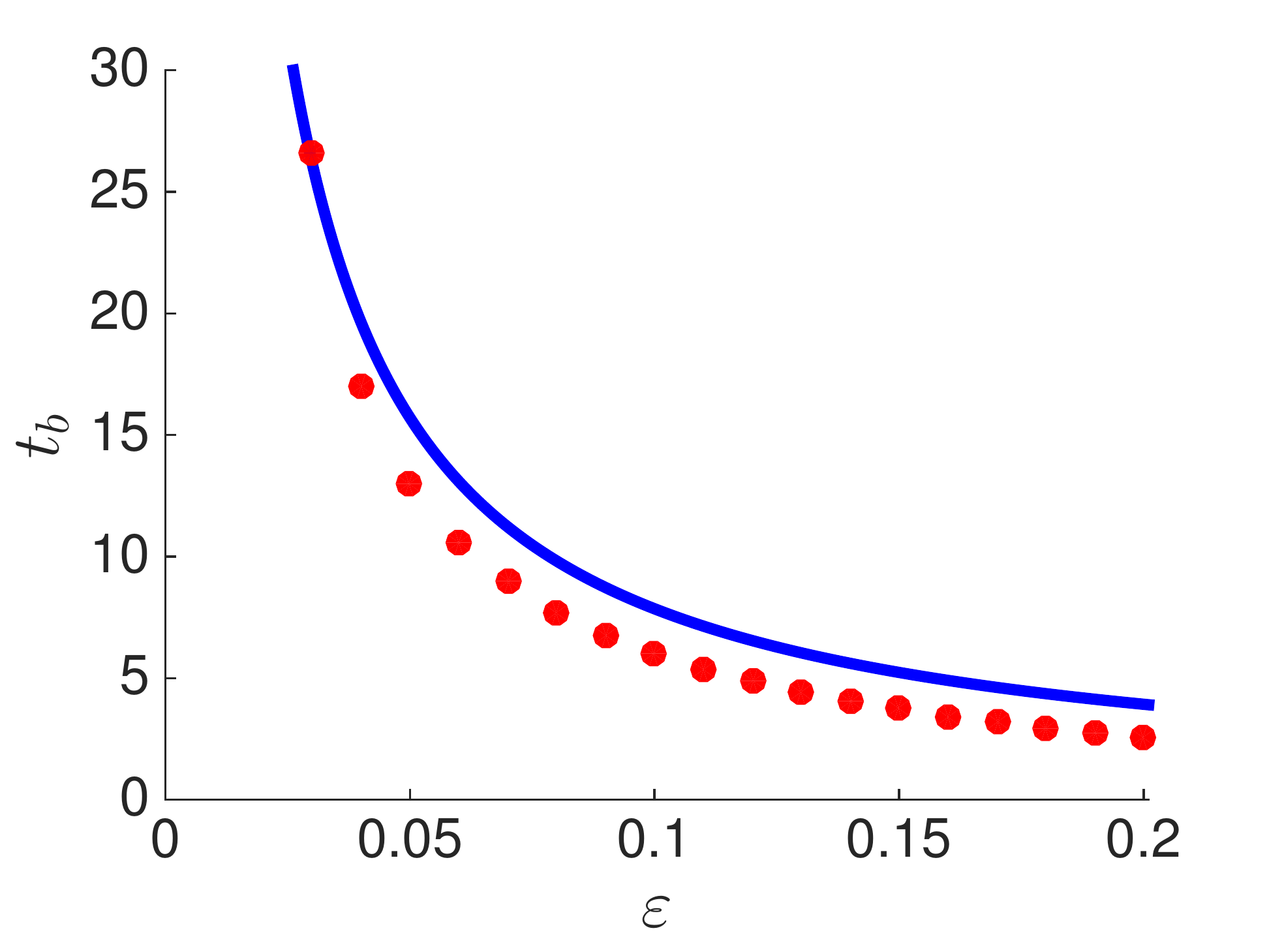}\end{center}
\caption{Slow passage of a bump on $x \in [- \pi, \pi]$ for a cosine weight function $w(x) = \cos (x)$. ({\bf A}) Amplitude of the even mode $A_e(t)$ slowly decreases with time. Numerical simulations (dashed lines) of (\ref{bfield}) are compared to the trajectory $\sqrt{2}(1 - \ve \tan (\ve t))$ determined by theory (solid lines). ({\bf B}) The duration of bottleneck increases as the distance to the bifurcation is decreased. Simulations (dots) are well fit by the theory $t_b = \pi / [4 \ve]$ (solid line).}
\label{fig3}
\end{figure}

Lastly, note when $\mu >0$, there are no fixed points of the differential equation (\ref{sna2}). However, starting at the initial condition $A_e(0) = 0$ (correspondingly $u(x,0) = U_c(x)$), we find that the dynamics of the amplitude $A_e(\tau)$ are strongly determined by the {\em ghost} of the fixed point at $A_e = 0$ \cite{strogatzbook}. Note in Fig. \ref{fig2}{\bf A} that the transient bump retains a shape much like that of the critical bump for an appreciable period of time before extinguishing. Trajectories of the full system (\ref{bfield}) evolve more slowly when the distance to the bifurcation $|\theta - \theta_{c}| = |\mu| \ve^2$ is smaller. Solving for $A_e(\tau)$ in this specific case and reverting the the original time coordinate $t = \tau/ \varepsilon$, we find
\begin{align}
A_e(t) = -\frac{w(0) \sqrt{\mu}}{\sqrt{|w'(2a_c)|}} \tan \left( \varepsilon \sqrt{\mu \cdot |w'(2a_c)|} t/ w(0) \right). \label{Aetan}
\end{align} 
Thus, the residence time $t_b$ in the {\em bottleneck}, or neighborhood of the {\em ghost} of the fixed point $A_e=0$, is given by the amount of time it takes for $A_e(t)$ to traverse to some set value. Of course, this is dependent on the bifurcation parameter $\mu$. For illustration, we examine how long it takes until $A_e(t_b) = -1$. By explicitly focusing on the region where $|A_e(t_b)| \leq 1$, we are roughly restricting to the time interval during which $|u(x,t) - U_c(x)| = {\mc O}(\ve)$, where we would expect the expansion (\ref{banz}) to be valid. Using the formula (\ref{Aetan}), it is straightforward to find that
\begin{align}
t_b = \frac{w(0)}{\varepsilon \sqrt{\mu \cdot |w'(2a_c)|} } \tan^{-1} \left( \frac{\sqrt{|w'(2a_c)|}}{w(0) \sqrt{\mu}} \right).  \label{btime}
\end{align}
We compare this formula to the results of numerical simulations in Fig. \ref{fig2}{\bf B}, utilizing the difference of Gaussians weight function $w(x) = \e^{-x^2} - A \e^{-x^2/\sigma^2}$ on $x \in (- \infty, \infty)$. Comparisons are made by noting that when $A_e(t_b) = -1$, then $u(x,t) \approx U_c(x) - \ve \psi_e(x)$, so that the peak of the activity profile will be
\begin{align*}
u(0,t_b) \approx U_c(0) - \ve \psi_e(0) = W(a_c) - W(-a_c) - \frac{2 w(a_c) \ve}{w(0)}.
\end{align*}
Notice in Fig. \ref{fig2}{\bf C},{\bf D} that, as predicted, the time spent in the bottleneck increases as the amplitude of the small parameter $\ve$ is decreased. The attracting impact of the ghost is stronger when the parameters of the system lie closer to the bottleneck. For further comparison, we consider the case $w(x) = \cos (x)$ in Fig. \ref{fig3}. In this case the constituent functions $a_c = \pi /4$, $w(0) = 1$, and $w(2a_c) = -1$. Furthermore, by setting $\mu = 1$ the formulas for the amplitude (\ref{Aetan}) and residence time (\ref{btime}) simplify considerably to $A_e(t) = - \tan (\ve t)$ and $t_b = \pi/ [4 \ve]$.

\subsection{Amplitude equations for smooth nonlinearities} Our nonlinear analysis in the case of Heaviside nonlinearities $f(u) \equiv H(u - \theta)$ made extensive use of the specific form of the distributional derivatives. Inner products with these functions lead to dynamical equations focused on a finite number of discrete points in space, rather than over the spatial continuum $x \in \Omega$. Here, we show it is straightforward to extend this analysis to the case of arbitrary smooth nonlinearities $f(u)$. There are several detailed analyses of stationary bumps in neural field with smooth firing rate, showing a similar bifurcation structure to that presented in Fig. \ref{fig1}: a stable and an unstable branch of bump solutions annihilate in a saddle-node bifurcation as the threshold of the firing rate function is increased. We refrain from such a detailed analysis here and refer the reader to these works \cite{kishimoto79,laing02,faugeras09,coombes10,veltz10,kilpatrick13}. Again, defining $\theta = \theta_c + \mu \ve^2$, $\ve \ll 1$, so $\mu$ determines the distance of $\theta$ from the bifurcation and on which side of $\theta_c$ it lies. Following our previous analysis, we utilize the ansatz (\ref{banz}) and rescale time $\tau = \ve \tau$. In this case, $\psi_o(x)$ and $\psi_e(x)$ will still be odd and even eigenmodes associated with the linear stability of stationary bump solutions to (\ref{bfield}). At the criticality $\theta \equiv \theta_c$, their associated eigenvalues will be $\lambda_o = \lambda_e\equiv 0$, as in the case of Heaviside firing rates \cite{veltz10}. Expanding (\ref{bfield}) in orders of $\ve$ using the ansatz (\ref{banz}) yields a similar amplitude equation to (\ref{snamp}) at ${\mc O}(\ve^2)$. Again, we apply solvability conditions to the equation for $u_2$.
After canceling odd terms and isolating the derivatives $A_j'$, we find the amplitudes $A_j$ satisfy the system:
\begin{subequations} \label{smoothampe}
\begin{align}
\frac{\d A_o}{\d t} &= \frac{\langle \varphi_o, w* \left[ f''(U_c) \psi_e^2 \right] \rangle}{2 \langle \varphi_o , \psi_o \rangle} A_e( \tau)^2, \label{smoothampa} \\
\frac{\d A_e}{\d \tau} & = - \mu \frac{\langle \varphi_e, w*f'(U_c) \rangle}{\langle \varphi_e, \psi_e \rangle} + \frac{\langle \varphi_e, w*\left[ f''(U_c) \psi_e^2 \right] \rangle}{2 \langle \varphi_e, \psi_e \rangle} A_e(\tau)^2.  \label{smoothampb}
\end{align}
\end{subequations}

We can derive the coefficients in the system (\ref{smoothampe}) by computing the inner products therein. To do so, we must choose a specific nonlinearity, such as the sigmoid (\ref{sig}), and a weight kernel. For illustration, we consider the cosine kernel $w(x) = \cos (x)$ on the ring $x \in \Omega = [- \pi, \pi]$ with periodic boundaries. As shown in previous studies, the bump solution $U_c(x) = A_c \cos x$ while the eigenmodes $\psi_o(x) = \sin (x)$ and $\psi_e(x) = \cos (x)$ \cite{hansel98,veltz10,kilpatrick13}. Since ${\mc L} \psi_j \equiv 0$ for $j=o,e$, this means
\begin{align*}
\sin (x) = \int_{- \pi}^{\pi} \cos (x-y) f'(A_c \cos (y)) \sin (y) \d y = \sin x \int_{- \pi}^{\pi} \sin^2(y) f'(A_c \cos y) \d y,
\end{align*}
where we have used $\cos (x-y) = \cos x \cos y + \sin x \sin y$, and
\begin{align*}
\cos (x) = \int_{- \pi}^{\pi} \cos (x-y) f'(A_c \cos (y)) \cos (y) \d y = \cos x \int_{- \pi}^{\pi} \cos^2(y) f'(A_c \cos y) \d y,
\end{align*}
so that we can write
\begin{align}
\int_{- \pi}^{\pi} \sin^2(y) f'(A_c \cos y) \d y \equiv 1, \hspace{5mm} \int_{- \pi}^{\pi} \cos^2(y) f'(A_c \cos y) \d y \equiv 1. \label{emodeid}
\end{align}
The identities (\ref{emodeid}) allow us to compute
\begin{align*}
\langle \varphi_o, \psi_o \rangle = \int_{- \pi}^{\pi} f'(A_c \cos (y)) \sin (y)^2 \d y = 1,
\end{align*}
and
\begin{align*}
\langle \varphi_e, \psi_e \rangle = \int_{- \pi}^{\pi} f'(A_c \sin (y)) \cos (y)^2 \d y = 1.
\end{align*}
Furthermore,
\begin{align}
\langle \varphi_o, w* \left[ f''(U_c) \psi_e^2 \right] \rangle &= \int_{- \pi}^{\pi}f''(U_c(y)) \psi_e(y)^2 \int_{- \pi}^{\pi} \cos (x-y) f'(A_c \cos (y)) \sin (y) \d x \d y \nonumber \\
&= \int_{- \pi}^{\pi}f''(U_c(y)) \cos(y)^2 \sin (y) \d y = 0, \label{smoddcoeff}
\end{align}
where the last equality holds due to the integrand being odd. Thus, the equation (\ref{smoothampa}) reduces to $A_o'(t) = 0$, so $A_o(t) \equiv \bar{A}_o$. Now, we can calculate the coefficients of the $A_e$ amplitude equation. First by utilizing the fact that $\int_{- \pi}^{\pi} w(x-y) \varphi_e(y) \d y = \psi_e(x)$, we can compute
\begin{align}
\langle \varphi_e, w* f'(U_c) \rangle = \int_{- \pi}^{\pi} f'(A_c \cos (x) ) \cos (x) \d x = \langle \varphi_e, 1 \rangle.
\end{align}
Lastly, we can simplify the integrals in the quadratic term by again making use of the identity $\int_{- \pi}^{\pi} w(x-y) \varphi_e(y) \d y = \psi_e(x)$, so
\begin{align}
\langle \varphi_e, w*\left[f''(U_c) \psi_e^2\right] \rangle= \int_{- \pi}^{\pi} f''(A_c \cos (x)) \cos^3(x) \d x = \langle f''(U_c), \psi_e^3 \rangle.
\end{align}
so we can simplify (\ref{smoothampb}) to
\begin{align}
\frac{\d A_e}{\d \tau} = - \mu \langle \varphi_e, 1 \rangle + \frac{1}{2} \langle f''(U_c), \psi_e^3 \rangle A_e(\tau)^2.
\end{align}

\section{Stochastic neural fields near the saddle-node}
\label{stochsys}
We now study the impact of stochastic forcing near the saddle-node bifurcation of bumps. Our analysis utilizes the spatially extended Langevin equation with additive noise (\ref{nfield}). Guided by our analysis of the deterministic system (\ref{bfield}), we will utilize an expansion in the small parameter $\ve$, which determines the distance of the system from the saddle-node. To formally derive stochastic amplitude equations, we must specify the scaling of the noise amplitude $\epsilon$ as it relates to the small parameter $\ve$, as this will determine the level of the perturbation hierarchy wherein the noise term $\d W$ will appear. We opt for the scaling $\epsilon = \ve^{5/2}$, as this introduces a nontrivial interaction between the nonlinear amplitude equation for $A_e$ and the noise.

It is important to note that our derivations are only carried up to ${\mc O}(\ve^2)$ in the hierarchy of the regular perturbation expansion in $\ve$. Were we to continue this expansion further, we would likely find that the $\epsilon = \ve^{5/2}$ noise term does indeed shift the location of the bifurcation at higher order as in  \cite{blomker07,hutt08}. Thus, as the amplitude of noise is increased, the validity of the expansion we derive here will begin to break down, since the terms beyond ${\mc O}(\ve^2)$ will have a more substantial effect on the dynamics. Hence, the results we derive in this section are valid for small noise levels only. An understanding of the effects of larger noise terms, employing scalings $\epsilon = \ve^p$ with $p<5/2$, warrants further study which is beyond the scope of our current work.

\subsection{Stochastic amplitude equation for bumps} Motivated by our quantitative analysis in the noise-free case, we rescale time in the stochastic term of (\ref{nfield}) using $\tau = \ve t$, so
\begin{align}
\d u (x,t) = \left[ - u(x,t) + \int_{\Omega} w(x-y) f(u(y,t)) \d y \right] \d t + \ve^2 \d \hat{W}(x, \tau),  \label{sfield}
\end{align}
where $\d \hat{W}(x,\tau) : = \sqrt{\ve} \d W(x, \ve^{-1}\tau)$ is a rescaled version of the Wiener process $\d W$ that is independent of $\ve$ \cite{gardiner04}. We then apply the ansatz (\ref{banz}) once again and take Heaviside firing rate functions (\ref{H}), thus finding (\ref{ordepssum}) at ${\mc O}(\ve)$. The ${\mc O}(\ve)$ equation is satisfied due to the fact that $\psi_e \in {\mc N}({\mc L})$, where ${\mc L}$ is the linear operator given by (\ref{linop}). Finally, proceeding to ${\mc O}( \ve^2 )$, we find
\begin{align}
{\mc L}\left[ A_o \psi_o + u_2 \right] \d t = & \d A_e \psi_e + \d A_o \psi_o + \mu \int_{\Omega} w(x-y) H'(U_c(y) - \theta_c) \d y \d t \label{oep2noise} \\
& - \frac{A_e^2}{2} \int_{\Omega} w(x-y) H''(U_c(y) - \theta_c) \psi_e(y)^2 \d y \d t + \d \hat{W}. \nonumber
\end{align}
As before, the $\psi_o$ terms on the left vanish since ${\mc L}\psi_o \equiv 0$, and we ensure a bounded solution to (\ref{oep2noise}) exists by requiring the inhomogeneous part is orthogonal to $\varphi_o, \varphi_e \in {\mc N}({\mc L}^*)$, where ${\mc L}^*$ is the adjoint linear operator given by (\ref{adjop}). Taking inner products yields
\begin{align}
0 = & \left\langle \varphi_j, \d A_e(\tau) \psi_e(x) + \d A_o(t) \psi_o(x) + \mu w*H'(U_c - \theta_c) \d t \right. \label{noiseip} \\
& \left. - \frac{A_e(\tau)^2}{2} w* \left[ H''(U_c - \theta_c) \psi_e^2 \right] \d t + \d \hat{W} \right\rangle, \nonumber
\end{align}
for $j=o,e$. Isolating temporal derivatives, we find the amplitudes $A_o(t)$ and $A_e(\tau)$ obey the following pair of nonlinear stochastic differential equations
\begin{subequations} \label{stochamp1}
\begin{align}
\d A_o(t) =& \frac{\langle \varphi_o, w*\left[ H''(U_c - \theta_c) \psi_e^2 \right] \rangle}{2 \langle \varphi_o, \psi_o \rangle }A_e(\tau)^2 \d t - \frac{\langle \varphi_o , \d \hat{W} \rangle}{\langle \varphi_o, \psi_o \rangle} \\
\d A_e(\tau) =& - \mu \frac{\langle \varphi_e, w*\left[ H'(U_c - \theta_c) \right] \rangle}{\langle \varphi_e, \psi_e \rangle} + \frac{\langle \varphi_e, w* \left[ H''(U_c - \theta_c) \psi_e^2 \right] \rangle}{2 \langle \varphi_e, \psi_e \rangle} A_e(\tau)^2 \\
& - \frac{\langle \varphi_e , \d \hat{W} \rangle}{\langle \varphi_e, \psi_e \rangle}. \nonumber
\end{align}
\end{subequations}
Utilizing the formulas for $H'(U_c - \theta_c)$ (\ref{Hdiff}) and $H''(U_c - \theta_c)$ (\ref{H2diff1}) we derived in the previous section, we can simplify the expressions in (\ref{stochamp1}). Additionally, we make use of the fact that
\begin{align*}
\d \hat{\mc W}_o(\tau) : = - \frac{\langle \varphi_o, \d \hat{W} \rangle}{\langle \varphi_o, \psi_o \rangle} &= -\frac{1}{2} \left[ \psi_o(-a_c) \d \hat{W}(-a_c,\tau) + \psi_o(a_c) \d \hat{W}(a_c,\tau) \right] \\
&= \frac{\d \hat{W}(-a_c,\tau) - \d \hat{W}(a_c,\tau)}{2}, \\
\d \hat{\mc W}_e(\tau) : = - \frac{\langle \varphi_e, \d \hat{W} \rangle}{\langle \varphi_e, \psi_e \rangle} &= -\frac{1}{2} \left[ \psi_e(-a_c) \d \hat{W}(-a_c,\tau) + \psi_e(a_c) \d \hat{W}(a_c,\tau) \right] \\
&= -\frac{\d \hat{W}(a_c,\tau) + \d \hat{W}(-a_c,\tau)}{2}. \\
\end{align*}
Utilizing the fact that $\langle \d \hat{W}(x,\tau) \d \hat{W}(y,\tau') \rangle = C(x-y) \delta( \tau - \tau') \d \tau \d \tau'$, it is straightforward to compute the variances $\langle \hat{\mc W}_o(\tau)^2 \rangle = D_o \tau = (C(0)-C(2a_c)) \tau /2$ and  $\langle \hat{\mc W}_e(\tau)^2 \rangle = D_e \tau = (C(0)+C(2a_c)) \tau /2$. Clearly, for spatially flat correlation functions $C(x) \equiv \bar{C}$, noise will have no impact on the odd amplitude $A_o(t)$ since $D_o \equiv 0$. Thus, (\ref{stochamp1}) becomes
\begin{subequations} \label{stochamp2}
\begin{align}
\d A_o(t) &= \sqrt{\ve} \d {\mc W}_o(t), \label{saodd} \\
\d A_e(\tau) &= - \mu \d \tau - \frac{|w'(2a_c)|}{w(0)^2} A_e(\tau)^2 \d \tau + \d \hat{\mc W}_e(\tau), \label{saeven}
\end{align}
\end{subequations}
where we have converted the noise term in (\ref{saodd}) back to the original time coordinate: $\d {\mc W}_o(t) = \d \hat{\mc W}_o(\ve t)/ \sqrt{\ve}$ \cite{gardiner04}. Note that in equation (\ref{saodd}), we essentially recover the diffusion approximation of the translating mode of the bump $\langle A_o(t)^2 \rangle = \ve D_ot$, which is analyzed in \cite{kilpatrick13}. Equation (\ref{saeven}) is a stochastic amplitude equation, so that the noise term $\d {\mc W}_e$ is projected onto the direction of the neutrally stable even perturbation $\psi_e$.

\begin{figure}

\begin{center} \includegraphics[width=6cm]{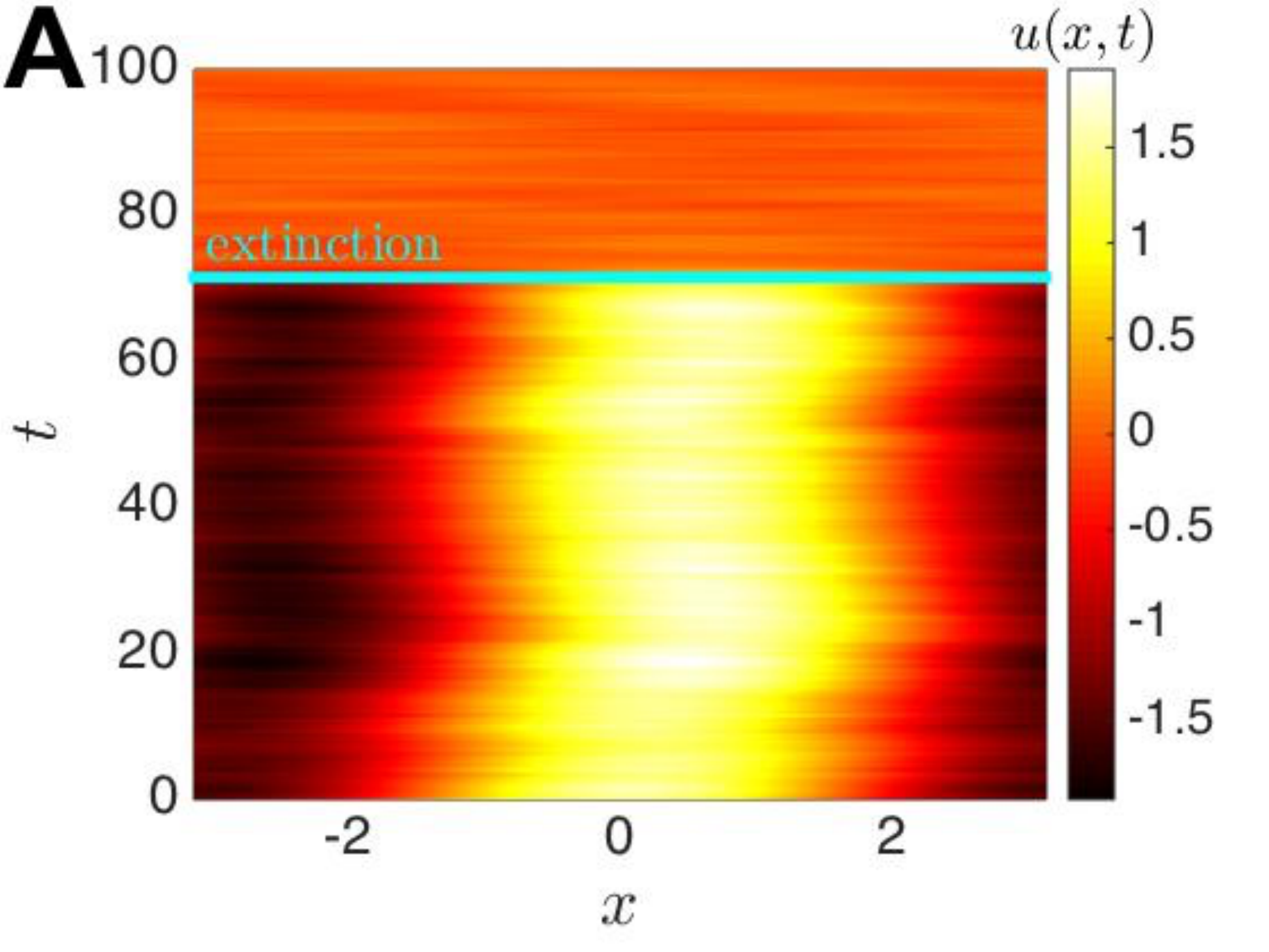} \includegraphics[width=6cm]{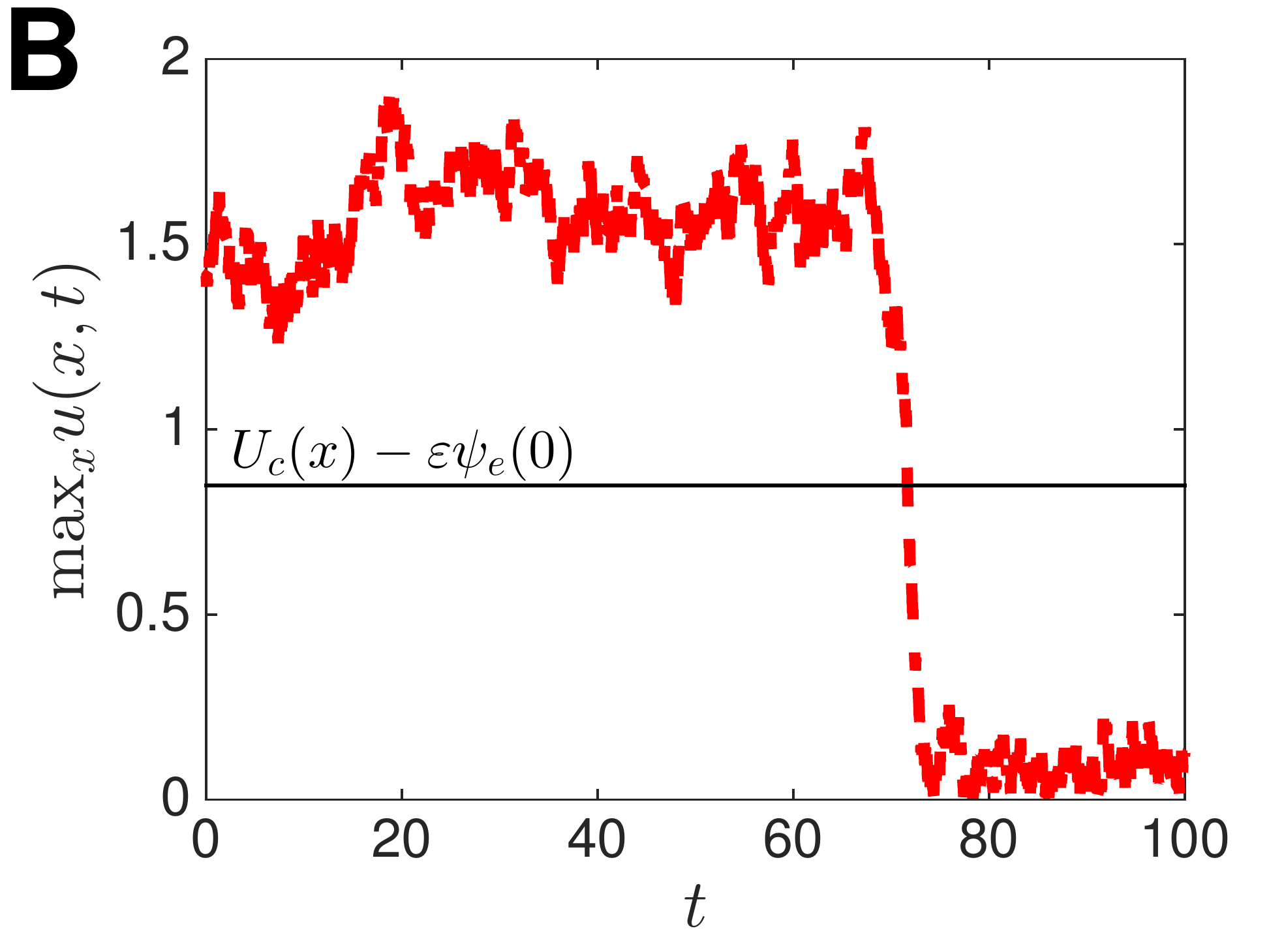} \end{center}
\caption{Noise-induced extinction of bumps in the stochastic neural field (\ref{nfield}) on $x\in [- \pi, \pi]$ for a cosine weight $w(x) = \cos (x)$. ({\bf A}) A single realization of the equation (\ref{nfield}) with the initial condition $u(x,0) = U_c(x) = \sqrt{2} \cos (x)$ leads to a stochastically wandering bump that eventually crosses a separatrix at $t \approx 70$, leading to extinction. The noise-free system possesses a stable bump solution since $\mu = -0.2<0$; $\ve = 0.4$. ({\bf B}) The large deviation can easily be detected by tracking ${\rm max}_x u(x,t)$, which departs the bottleneck of the noise-free system, whose lower bound lies at ${\rm max}_x \left[ U_c(x) - \ve \psi_e(0) \right] = \sqrt{2}(1- \ve)$.}
\label{fig4}
\end{figure}

\subsection{Metastability and bump extinction} 
\begin{figure}
\begin{center} \includegraphics[width=6cm]{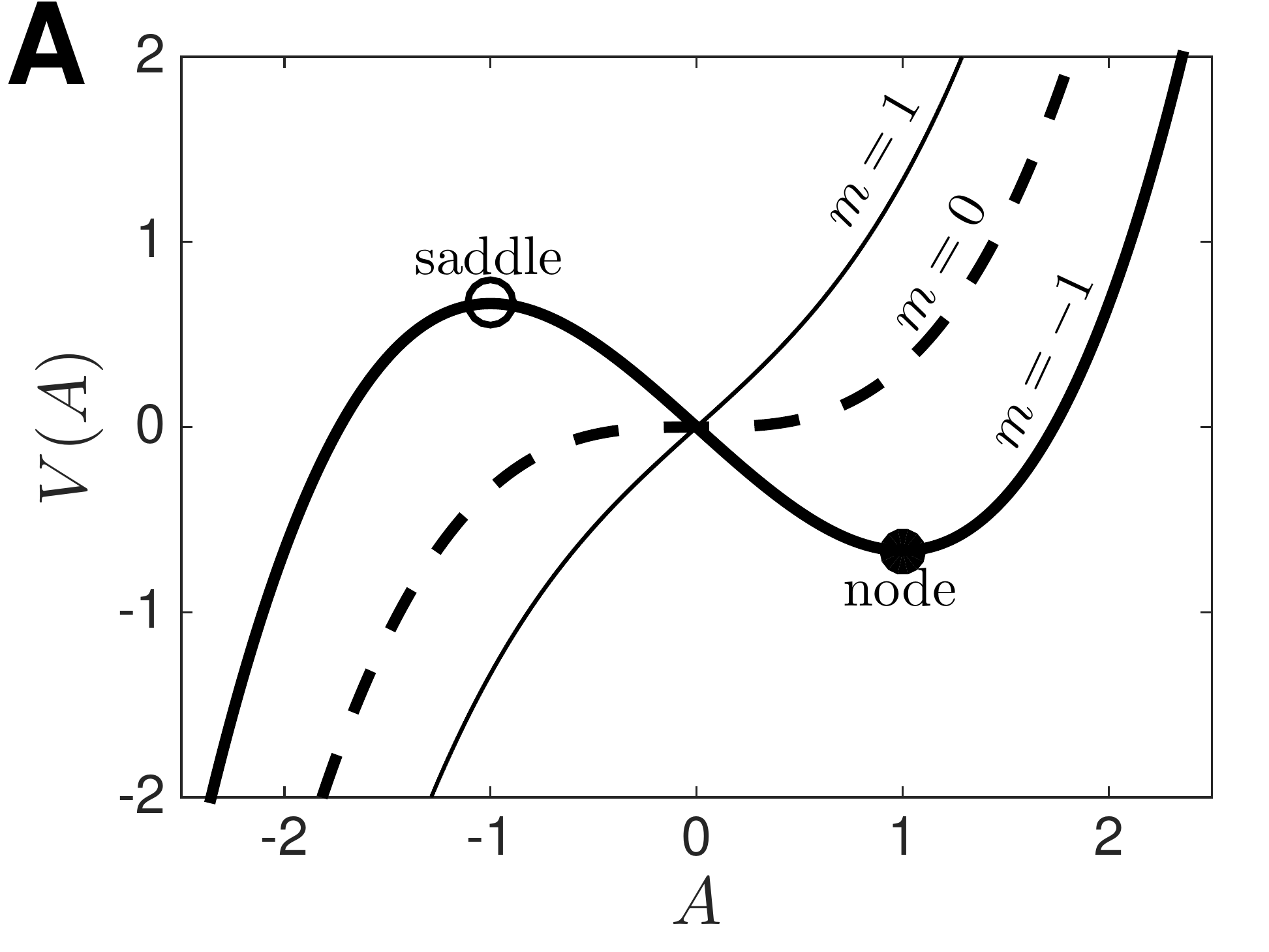} \includegraphics[width=6cm]{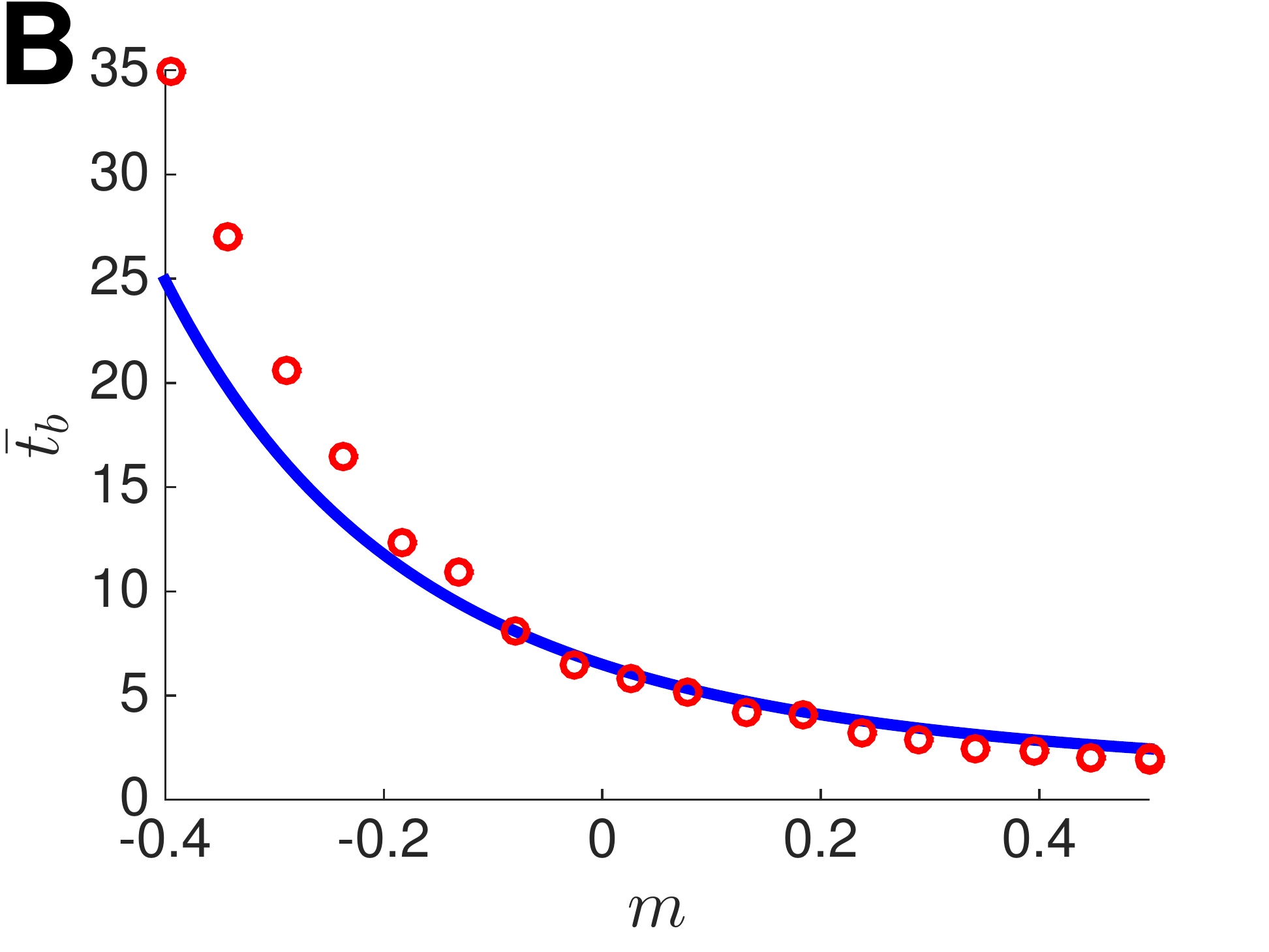} \end{center}
\caption{({\bf A}) Potential function (\ref{snpot}) associated with the stochastic amplitude equation (\ref{sarescale}) has zero ($m>0$); one ($m\equiv 0$); or two ($m<0$) extrema - associated with equilibria of $\dot{A} = -m - A^2$. When $m<0$, crossing the saddle point requires stochastic forcing. ({\bf B}) Mean time $\bar{t}_b$ until bump extinction is approximated by a mean first passage time problem of the stochastic amplitude equation (\ref{sarescale}). Numerical simulations (circles) of the full system (\ref{nfield}) are well approximated by this theory (line) given by (\ref{mfpt3}) for $\ve = 0.6$.}
\label{fig5}
\end{figure}

To analyze the one-dimensional nonlinear SDE (\ref{saeven}), we further rescale the equation by setting $A : = \frac{|w'(2a_c)|}{w(0)^2} A_e$:
\begin{align}
\d A(\tau) = - \left[ m + A(\tau)^2 \right] \d t + \d \hat{\mc W}(\tau),  \label{sarescale}
\end{align}
where $m := \frac{|w'(2a_c)|}{w(0)^2} \mu$. Thus, the effective diffusion coefficient of the rescaled noise term is $\langle \hat{\mc W}(\tau)^2 \rangle = D \tau = w'(2a_c)^2 (C(0) + C(2a_c)) \tau /\left[ 2 w(0)^4 \right]$. Note the rescaled equation (\ref{sarescale}) has an effective potential \cite{lindner03,strogatzbook}:
\begin{align}
V(A) = \frac{A^3}{3} + m A ,  \label{snpot}
\end{align}
the derivative $V'(A)$ of which yields the deterministic part of the right hand side. As the bifurcation parameter $m$ is varied, the potential exhibits a minimum (at $A=\sqrt{m}$) and a maximum (at $A=- \sqrt{m}$) when $m<0$, a saddle point (at $A = 0$) when $m=0$, and no extrema for $m>0$ (Fig. \ref{fig5}{\bf A}). For all parameter values $m$, the state of the stochastic system (\ref{sarescale}) will eventually escape to the limit $A \to - \infty$ as $\tau \to \infty$. Such trajectories were observed in the noise-free system in the case $m>0$, as demonstrated in Fig. \ref{fig2} of the previous section. However, we show here that noise qualitatively alters the dynamics of the system, so its state will not remain in the vicinity of the stable attractor (at $A=\sqrt{m}$) when $m<0$. 

As before, we study the problem of bump extinction using the stochastic amplitude equation (\ref{sarescale}) in the case $m>0$. We show that the noise decreases the average amount of time until an extinction event will occur. For clarity, we assume the initial condition $A(0) = 0$ (correspondingly $u(x,0) = U_c(x)$). We take the bottleneck to be the region $A_e \in [ -1,1]$, which in the rescaled variable is $A \in [-|w'(2a_c)|/w(0)^2, |w'(2a_c)|/w(0)^2]$. The residence time $\tau_b$ in the bottleneck is given by the amount of time it takes for $A$ to escape this region. We can determine the statistics of $\tau_b$ by considering it as a first passage time problem. 

Let $p(A,\tau)$ be the probability density for the stochastic process $A(\tau)$ given the initial condition $A(0) = A_0$. Then the corresponding Fokker-Planck equation is given
\begin{align}
\frac{\pd p}{\pd \tau} = \frac{\pd \left[ (m+A^2) p(A, \tau) \right]}{\pd A} + \frac{D}{2} \frac{\pd^2 p(A,\tau)}{\pd A^2} \equiv - \frac{\pd J (A,\tau)}{\pd A},  \label{fokplanck}
\end{align}
where
\begin{align}
J(A,\tau) = - \frac{D}{2} \frac{\pd p(A,\tau)}{\pd A} - (m + A^2) p(A,\tau), \label{flux}
\end{align}
and $p(A,0) = \delta (A - A_0)$. We focus on the three different scenarios discussed above. First, if $m<0$, there there is a single stable fixed point of the deterministic equation $\dot{A} = - m - A^2$ at $A = \sqrt{m}$ and a single unstable fixed point at $A = - \sqrt{m}$. The basin of attraction of $A = \sqrt{m}$ is given by the interval $(- \sqrt{m} , \infty)$. When $D>0$, fluctuations can induce rare transitions on exponentially long timescales whereby $A(\tau)$ crosses the point $A = - \sqrt{m}$, leaving the basin of attraction. For the non-generic case $m = 0$, the timescale of departure scales algebraically \cite{sigeti89}. When $m>0$, noise simply modulates the flows of the deterministic equation $\dot{A} = -m - A^2$, leading to an average speed-up in the departure from the bottleneck. In general, we consider solving the first passage time problem as an escape from the domain $(- \alpha, \infty)$ where $\alpha := \frac{|w'(2a_c)|}{w(0)^2}$ (equivalently where $A_e = -1$) \cite{gardiner04}. To do so, we impose an absorbing boundary condition at $- \alpha$: $p(-\alpha, \tau) = 0$. Now let $T(A)$ denote the stochastic first passage time for which (\ref{sarescale}) first reaches the point $- \alpha$, given it started at $A \in (- \alpha, \infty)$. The first passage time distribution is related to the survival probability that the system has not yet reached $-\alpha$:
\begin{align*}
S(\tau) \equiv \int_{- \alpha}^{\infty} p(A,\tau) \d A,
\end{align*}
which is $S(\tau) : = {\rm Pr}(t>T(A))$, so the first passage time density is \cite{gardiner04}
\begin{align*}
F(\tau) = - \frac{\d S}{\d \tau} = - \int_{-\alpha}^{\infty} \frac{\pd p}{\pd \tau}(A,\tau) \d A.
\end{align*}
Substituting for the expression for $\pd p/ \pd \tau$ using the Fokker-Planck equation (\ref{fokplanck}) and the formula for the flux (\ref{flux}) shows
\begin{align*}
F(\tau) = \int_{- \alpha}^{\infty} \frac{\pd J(A,\tau)}{\pd A} \d A = - J(-\alpha, \tau),
\end{align*}
where we have utilized the fact that $\lim_{A \to \infty} J(A,\tau) = 0$. Thus, the first passage time density $F(\tau)$ can be interpreted as the total probability flux through the absorbing boundary at $A = - \alpha$. To calculate the mean first passage time ${\mc T}(A) := \langle T(A) \rangle$, we use standard analysis to associate ${\mc T}(A)$ with the solution of the backward equation \cite{gardiner04}:
\begin{align}
-(m + A^2) \frac{\d {\mc T}}{\d A} + \frac{D}{2} \frac{\d^2 {\mc T}}{\d A^2} = - 1,  \label{mfpt1}
\end{align}
with the boundary conditions ${\mc T}(- \alpha) = 0$ and ${\mc T}'(\infty) = 0$. Solving (\ref{mfpt1}) yields the closed form solution
\begin{align}
{\mc T}(A) = \frac{2}{D} \int_{-\alpha}^{A} \int_{y}^{\infty} \frac{\phi (z)}{\phi(y)} \d z \d y,  \label{mfpt2}
\end{align}
where
\begin{align*}
\phi (A) = \exp \left[ \frac{2\left[ V(- \alpha) - V(A) \right]}{D}\right],
\end{align*}
and $V(x)$ is the potential function (\ref{snpot}). Explicit expressions for the integral (\ref{mfpt2}) can be found in some special cases \cite{sigeti89,lindner03}. For our purposes, we simply integrate (\ref{mfpt2}) numerically to generate theoretical relationships between the mean first passage time and model parameters. For comparison, we focus on the case the weight function $w(x) = \cos (x)$ and the correlations $C(x) = \cos (x)$, so that $U_c(x) = \sqrt{2} \cos (x)$, $a_c = \frac{\pi}{4}$, $w(0) = 1$, $w'(2a_c) = -1$, $C(0) = 1$, and $C(2a_c) = 0$. Therefore, $\alpha = 1$, $m = \mu$, $D = 1/2$. This allows us to write the formula (\ref{mfpt2}) at $A = 0$ as
\begin{align}
{\mc T}(0) = 4 \int_{-1}^{0} \int_{y}^{\infty} \exp \left[ 4 \left( \frac{z^3-y^3}{3} + \mu (z-y) \right) \right]  \d z \d y.  \label{mfpt3}
\end{align}
Lastly, note that by rescaling time $t = \ve \tau$, we have that the mean first passage time in units of $t$ will be $\bar{t}_b = {\mc T}(0)/\ve$. We compare our theory (\ref{mfpt3}) with the results of numerical simulations of the full stochastic neural field (\ref{nfield}) in Fig. \ref{fig5}{\bf B}. Note there is some discrepancy between our numerical simulations and theory as $m$ is decreased. One of the primary reasons for this deviation is likely because of the moderate level of noise ($\varepsilon = 0.6$) used in comparison to the small parameter assumption ($\varepsilon \ll 1$) using in the theory we have developed. Any minor mismatch will be exacerbated by the fact that mean first passage times for escape problems depend exponentially on parameters like noise amplitude and well depth, as in (\ref{mfpt3}). Nonetheless, the theory does provide a rough estimate of the mean first passage times for smaller values of the parameter $m$.

\section{Discussion} We have developed a weakly nonlinear analysis for saddle-node bifurcations of bumps in deterministic and stochastic neural field equations. While most of our analysis has focused upon Heaviside firing rate functions, we have also demonstrated the techniques can easily be extended to arbitrary smooth nonlinearities. In the vicinity of the saddle-node, the dynamics of bump expansion/contraction can be described by a quadratic amplitude equation. For deterministic neural fields, this low dimensional approximation can be used to approximate the trajectory and lifetime of bumps as they slowly extinguish. To do so, we focused on the initial time epoch in the bottleneck surrounding the ghost of the critical bump $U_c(x)$. In stochastic neural fields with appropriate noise scaling, a stochastic amplitude equation for the even mode of the bump can be derived. Importantly, we must choose the noise amplitude to scale as $\epsilon = \ve^{5/2}$, in order for the noise term to appear in the stochastic version of the quadratic amplitude equation. We then cast the lifetime of the bump in terms of a mean first passage time problem of the reduced system, which is valid for the noise scaling we have chosen.

Our work extends a variety of recent studies that have derived low-dimensional nonlinear approximations of neural field pattern dynamics in the vicinity of bifurcations \cite{bressloff04,hutt08,folias11,kilpatrick13,kilpatrick14,bressloff15}. As in our work, most of these previous studies derived approximations where the location of the bifurcation was unaffected by noise terms. On the other hand, Hutt et al. showed that noise can in fact shift the position of Turing bifurcations in neural fields, and the amplitude of the bifurcation threshold shift was proportional to the noise variance \cite{hutt08}. Were we to have carried the hierarchy out to higher order, we would have found such a shift in the case we studied. Note, it was necessary in our work to apply a specific noise scaling ($\ve^{5/2}$), as compared to the distance from criticality ($\ve^2$), in order for the noise to simply appear as a modification of the even mode amplitude equation. Were we to have selected noise of larger amplitude, this could have induced bifurcation shifts at lower order, analogous to that found in \cite{hutt08}. Another potential future direction would be to consider the impact of axonal propagation delays \cite{hutt03} on the dynamics close to the saddle-node. As demonstrated in this work, the neural field (\ref{bfield}) is quite sensitive to small perturbations near criticality, so delays may alter the duration of the bottleneck or even shift the saddle-node bifurcation point. In previous work \cite{kilpatrick14}, we derived amplitude equations describing propagation-generating drift bifurcations that arise when linear adaptation is incorporated into (\ref{bfield}). We anticipate that similar analyses might be performed on networks with synaptic depression \cite{kilpatrick10}, although piecewise smooth methods may be necessary in the case of Heaviside firing rates \cite{kilpatrick10b}. Lastly, we note there have been recent efforts to systematically derive macroscopic descriptions of neural activity from recurrently coupled spiking networks \cite{buice13,laing14,montbrio15}. Such alternative descriptions can also capture the form of steady state solutions like stationary bumps \cite{laing14,qiu14}. Bumps in these models do exhibit saddle-node bifurcations similar to those observed in the Amari model \cite{amari77}. Thus, extending our methods to such models would merely require knowledge of the stationary equations defining bump solutions. Perturbation analysis along with solvability conditions could then yield the coefficients of the amplitude equations near the saddle-node.


\bibliographystyle{siam}
\bibliography{ghost}

\end{document}